\documentclass[prd,superscriptaddress,nofootinbib]{revtex4}

\usepackage[utf8]{inputenc}
\usepackage[english]{babel}
\usepackage{amsmath}
\usepackage{subfigure}
\usepackage{amsfonts}
\usepackage{amssymb}
\usepackage{graphicx}
\usepackage{color}

\begin{document}

\hfill{} CERN-TH-2022-022

\title{A Two-Component Dark Matter Model and \\ 
its Associated Gravitational Waves}

\author{Francesco Costa}
\email{francesco.costa@theorie.physik.uni-goettingen.de}
\affiliation{Institute for Theoretical Physics, 
Georg-August University G\"ottingen,
Friedrich-Hund-Platz 1, G\"ottingen, D-37077 Germany}
\author{Sarif Khan}
\thanks{Corresponding Author}
\email{sarif.khan@uni-goettingen.de}
\affiliation{Institute for Theoretical Physics, 
Georg-August University G\"ottingen,
Friedrich-Hund-Platz 1, G\"ottingen, D-37077 Germany}
\author{Jinsu Kim}
\thanks{Corresponding Author}
\email{jinsu.kim@cern.ch}
\affiliation{Theoretical Physics Department, CERN, 
1211 Geneva 23, Switzerland}

\begin{abstract} 
We consider an extension of the Standard Model that accounts for the muon $g-2$ tension and neutrino masses and study in detail dark matter phenomenology. The model under consideration includes a WIMP and a FIMP scalar dark matter candidates and thus gives rise to two-component dark matter scenarios. We discuss different regimes and mechanisms of production, including the novel freeze-in semi-production, and show that the WIMP and FIMP together compose the observed relic density today. The presence of the extra scalar fields allows phase transitions of the first order. We examine the evolution of the vacuum state and discuss stochastic gravitational wave signals associated with the first-order phase transition. We show that the gravitational wave signals may be probed by future gravitational wave experiments which may serve as a complementary detection signal.
\end{abstract}
\maketitle
%%%%%%%%%%%%%%%%%%%%%%%%%%%%%%%%%%%

%%%%%%%%%%%%%%%%%%%%%%%%%%%%%%%%%%%
\section{Introduction}
\label{sec:intro}
%%%%%%%%%%%%%%%%%%%%%%%%%%%%%%%%%%%

The Standard Model (SM) of particle physics proved to be very precise in describing the nature of the physical world. However, some of its problems were highlighted and studied in the past decades, including the neutrino masses, the existence of dark matter (DM), and the muon $g-2$ tension.
In the SM, neutrinos are massless. However, the evidence of neutrino oscillations indicates otherwise \cite{Super-Kamiokande:1998kpq,Gonzalez-Garcia:2002bkq}. The mass splitting from neutrino oscillation experiments is constrained to be $|\Delta m_{21}^{2}|= 7.42_{-0.20}^{+0.21} \times 10^{-5} \, {\rm eV}^{2}$ between the first and the second mass eigenstates, while it is $\Delta m_{32}^{2}=2.517_{-0.028}^{+0.026} \times 10^{-3} \, {\rm eV}^{2}$ for the second and the third \cite{Esteban:2020cvm}. Moreover, from cosmological data, we also have a bound on the sum of their masses $\sum_i m_{\nu_i} < 0.3 \, {\rm eV}$ \cite{Goobar:2006xz}. 

The recent data coming from Fermilab \cite{Muong-2:2021ojo} increased the tension between the SM theoretical prediction for the muon anomalous magnetic moment, the $g-2$ factor, and the experimental data. At the moment there is a $4.2 \sigma$ discrepancy,
\begin{align}
\Delta a_{\mu}=
a_{\mu}^{\rm exp}-a_{\mu}^{\rm SM}=
(2.51 \pm 0.59) \times 10^{-9}
\,,
\end{align}
suggesting the presence of new physics at a scale of hundreds of GeV.

Finally, the SM fails to accommodate one or more particles that may play the role of the DM. Ever since the proposal by Zwicky for a dark, collision-less, and matter-like component of the energy budget of the universe \cite{Zwicky:1933gu,Bertone:2016nfn}, evidences from different sources for a cold, particle-like DM have cumulated \cite{Ostriker:1973uit,Planck:2018vyg,Corbelli:1999af}. 
The most promising and studied solution to this problem is the Weekly Interacting Massive Particle (WIMP) \cite{Gunn:1978gr,Hut:1977zn,Lee:1977ua,Bertone:2004pz}. The WIMP DM is, however, strongly constrained by experimental data \cite{XENON:2018voc,CMS:2016lcl,MAGIC:2016xys, Arcadi:2017kky,PandaX-II:2016vec,LUX:2016ggv}. Thus, more attention has been drawn to alternative DM production mechanisms. For example, the freeze-in mechanism has gained increasing interest \cite{McDonald:2001vt,Choi:2005vq,Kusenko:2006rh,Hall:2009bx,Cheung:2011nn,Elahi:2014fsa,Arcadi:2015ffa,Bernal:2017kxu,Benakli:2017whb,Bernal:2018qlk,Bernal:2019mhf,Covi:2020pch,Khan:2020pso,Garcia:2020hyo,Bernal:2020qyu}. 
The abundance of a WIMP DM is produced through the freeze-out mechanism, which is a thermal process, and it is generally inversely proportional to the thermal cross section. Instead, the DM produced via the freeze-in mechanism, called Feebly Interacting Massive Particle (FIMP), is out of equilibrium with respect to the thermal bath of the SM particles. A small coupling between the visible sector and the DM is predicted, making this candidate more difficult to detect and to constrain with direct detection experiments.\footnote{
In Refs. \cite{Kim:2017mtc,Kim:2018xsp,Goudelis:2018xqi}, it was pointed out that such a small coupling can naturally arise in a clockwork framework \cite{Choi:2015fiu,Kaplan:2015fuy,Giudice:2016yja}.
}
Nonetheless, both freeze-out and freeze-in production mechanisms are physically viable and not mutually exclusive. It is thus worth exploring the possibility of multi-component DM scenarios, where both the WIMP and FIMP DM contribute to the current relic density $\Omega_{\rm DM} h^2 = 0.120 \pm 0.001$ observed by the Planck experiment \cite{Planck:2018vyg}. Recent studies on multi-component DM scenarios include Refs. \cite{Zurek:2008qg,Profumo:2009tb,Feldman:2010wy,Bian:2013wna,Biswas:2013nn,Bhattacharya:2013hva,Bian:2014cja,Belanger:2014vza,Esch:2014jpa,Arcadi:2016kmk,Bhattacharya:2016ysw,DuttaBanik:2016jzv,Bhattacharya:2017fid,Ahmed:2017dbb,Bernal:2018aon,Bhattacharya:2018cgx,Elahi:2019jeo,Borah:2019aeq,Bhattacharya:2019fgs,Yaguna:2019cvp,Abdallah:2019svm,Belanger:2020hyh,Choi:2021yps,DiazSaez:2021pfw,DiazSaez:2021pmg,Saez:2021qta,Belanger:2021lwd,Ho:2021ojb,Ho:2022erb,Bhattacharya:2022wtr,Das:2022oyx}.

In this paper, we consider an extension of the SM and explain the aforementioned three problems of the SM in a single unified framework.
A novel set-up is proposed where we introduce three massive right-handed (RH) neutrinos $N_R$ that, through the standard type-I seesaw mechanism \cite{Minkowski:1977sc,Gell-Mann:1979vob}, provide a mass to the SM neutrinos. The SM is then also extended with two SM-singlet scalar fields $\phi_1$ and $\phi_2$ that play the role of the WIMP-like DM and FIMP-like DM, respectively. 
Finally, we introduce an extra $U(1)_{L_{\mu}-L_ {\tau}}$ gauge symmetry with a related gauge boson $Z_{\mu \tau}$ which receives its mass from a second Higgs field $\phi_H$. The presence of a new massive gauge boson, with the vacuum expectation value (VEV) of $\phi_H$ around hundreds of GeV can solve the $g-2$ tension \cite{Abdallah:2011ew,Khalil:2015wua,Lindner:2016bgg,Chun:2016hzs,Calibbi:2018rzv,Arnan:2019uhr,Calibbi:2020emz,Athron:2021iuf}.

Appropriately assigning $U(1)_{L_{\mu}-L_ {\tau}}$ charges for the DM particles, we also address a novel production mechanism, namely the freeze-in mechanism by semi-production processes \cite{Bringmann:2021tjr,Hryczuk:2021qtz} like $\phi_1 \phi_2 \leftrightarrow \phi_2 \phi_2$, that is the inverse of the semi-annihilation process \cite{DEramo:2010keq}. This mechanism produces an exponentially increasing DM yield, and it typically requires a larger coupling than the standard freeze-in scenarios do, since the DM abundance is also suppressed by the small initial abundance which is generically required for the freeze-in production mechanism.

The evolution of the vacuum state of the scalar potential becomes non-trivial due to the three extra scalar fields. First-order phase transitions (FOPTs) may thus arise, producing stochastic gravitational wave (GW) signals \cite{Kamionkowski:1993fg} detectable by future GW experiments such as LISA \cite{Baker:2019nia} which is a space-based detector comprising of three spacecraft, utilising laser interferometry, DECIGO \cite{Seto:2001qf,Kawamura:2006up,Sato:2017dkf,Isoyama:2018rjb,Kawamura:2020pcg} which is a proposed GW antenna in space designed to observe GWs in the 0.1 -- 10 Hz frequency range, consisting of four clusters of LISA-like three spacecraft, and BBO \cite{Corbin:2005ny,Crowder:2005nr,Harry:2006fi} which is a proposed follow-up of the LISA experiment, aiming to form a triangular shape consisting of four LISA-like detectors, similar to the DECIGO. For recent studies on this subject, see, {\it e.g.}, Refs. \cite{Grojean:2006bp,Huber:2008hg,Espinosa:2008kw,Caprini:2015zlo,Artymowski:2016tme,Baldes:2017rcu,Beniwal:2018hyi,Hashino:2018zsi,Caprini:2018mtu,Bian:2018mkl,Bian:2018bxr,Bian:2019szo,Bian:2019kmg,Caprini:2019egz,Di:2020ivg,Zhou:2021cfu,Mohamadnejad:2021tke,Bian:2021dmp}. This possibility gives a complementary detection signal to the standard (in-)direct detection and collider searches that potentially can probe our model and unveil the nature of the DM. We present a region of the model parameter space that produces detectable GW signals from a FOPT, relieves the muon $g-2$ tension, gives masses to the SM neutrinos, and explains the correct DM abundance by a two-component WIMP-FIMP relic density.

The rest of the paper is organised as follows. We set up our model in Section \ref{sec:model}, introducing the particle content, the gauge groups, and the mass spectrum of the theory. We also present the standard type-I seesaw mechanism adopted to explain the neutrino masses, and we give a brief explanation of the muon $g-2$ tension. In Section \ref{sec:TCDM}, we discuss possible DM scenarios. We divide the parameter space into three regimes and study both one-component and two-component scenarios.
In Section \ref{sec:FOPTGW}, the FOPT and its associated GWs are studied. We showcase four benchmark points that explain the muon $g-2$, neutrino masses, and correct DM relic density. The benchmark points predict GW signals within the detectability of future GW experiments, in particular Ultimate-DECIGO, which is an ultimate, idealised version of the DECIGO, whose sensitivity is only limited by quantum noises. We conclude in Section \ref{sec:conc}.

%%%%%%%%%%%%%%%%%%%%%%%%%%%%%%%%%%%
\section{Model}
\label{sec:model}
%%%%%%%%%%%%%%%%%%%%%%%%%%%%%%%%%%%

We consider the following Lagrangian:
\begin{align}
\mathcal{L}=
\mathcal{L}_{\rm SM} + 
\mathcal{L}_{\phi_H} +
\mathcal{L}_{N} + 
\mathcal{L}_{\rm DM} + 
\mathcal{L}_{\rm int} -
\frac{1}{4} F_{\mu \tau}^{\alpha \beta} {F_{\mu \tau}}_{\alpha \beta}
\,, 
\label{eqn:lag}
\end{align}
which obeys the symmetry of the complete gauge group $SU(3)_c \times SU(2)_L \times U(1)_Y \times U(1)_{L_\mu - L_\tau}$, where $\mathcal{L}_{\rm SM}$ is the SM Lagrangian including the SM Higgs field $\phi_h$, $\mathcal{L}_{\phi_H}$ is the Lagrangian for the $U(1)_{L_\mu - L_\tau}$ Higgs field $\phi_H$,
\begin{align}
\mathcal{L}_{\phi_H} 
= (D_\mu\phi_H)^\dagger(D^\mu\phi_H) 
+ \mu_H^2|\phi_H|^2 - \lambda_H|\phi_H|^4
\,,
\end{align}
and $\mathcal{L}_{N}$ is the Lagrangian for the RH neutrinos containing their kinetic terms, mass terms, and Yukawa terms with the SM lepton doublets,
\begin{align}
\mathcal{L}_{N}&=
\sum_{i=e,\mu,\tau}\frac{i}{2}\bar{N_i}\gamma^{\mu}D^{N}_{\mu} N_{i} 
-\frac{1}{2}M_{ee}\bar{N_e^{c}}N_{e}
-\frac{1}{2}M_{\mu \tau}(\bar{N_{\mu}^{c}}N_{\tau}
+\bar{N_{\tau}^{c}}N_{\mu})
\nonumber \\
&\quad
-h_{e \mu}(\bar{N_{e}^{c}}N_{\mu} 
+\bar{N_{\mu}^{c}}N_{e})\phi_H^\dagger
- h_{e \tau}(\bar{N_{e}^{c}}N_{\tau} 
+ \bar{N_{\tau}^{c}}N_{e})\phi_H
-\sum_{i=e,\mu,\tau} y_{i} \bar{L_{i}}
\tilde{\phi}_{h} N_{i} +{\rm h.c.}
\,,
\label{eqn:lagN}
\end{align}
where $\tilde{\phi}_{h}=i\sigma_2\phi^*_h$, and $M_{ee}$ and $M_{\mu \tau}$ are constants whose mass-dimension is one, while $h_{e\mu}$, $h_{e \tau}$, and $y_i$ are dimensionless coupling constants. 
In Eq. \eqref{eqn:lag}, $\mathcal{L}_{\rm DM}$ is the DM Lagrangian that is given by
\begin{align}
\mathcal{L}_{\rm DM} =
\sum_{i=1,2} (D^{\mu}\phi_{i})^\dagger (D_{\mu}\phi_{i})
- \sum_{i = 1, 2}\mu_{i}^{2} \phi_{i}^{\dagger} \phi_{i} 
- \sum_{i = 1, 2}\lambda_{i} (\phi_{i}^{\dagger} \phi_{i})^{2}
-\lambda_{12} (\phi^{\dagger}_{1} \phi_1) (\phi^{\dagger}_{2} \phi_2) - \mu (\phi^{\dagger}_{1} \phi^{3}_{2} + {\rm h.c.})
\,.
\label{eqn:lagDM}
\end{align}
Furthermore, $\mathcal{L}_{\rm int}$ in Eq. \eqref{eqn:lag} contains all the interactions between the SM Higgs field $\phi_h$, the second $U(1)_{L_\mu - L_\tau}$ Higgs field $\phi_H$, and the DM fields $\phi_{1,2}$,
\begin{align}
\mathcal{L}_{\rm int} = 
- \lambda_{hH}(\phi_{h}^{\dagger} \phi_{h}) (\phi_{H}^{\dagger} \phi_{H}) 
- \sum_{i = 1,2 \; j= h, H}\lambda_{i j}(\phi_{i}^{\dagger} \phi_{i})
(\phi_{j}^{\dagger} \phi_{j}) 
\,.
\label{eqn:lagINT}
\end{align}
The covariant derivatives in Eqs. (\ref{eqn:lag}) -- (\ref{eqn:lagDM}) can generically be written as $D_{\nu} X = (\partial_{\nu} + i g_{\mu\tau} Q_{\mu \tau} (X) {Z_{\mu\tau}}_ \nu)X$, where $X$ is a SM-singlet field whose $U(1)_{L_\mu-L_\tau}$ charge is $Q_{\mu \tau} (X)$ (see Table \ref{tab:ParticleContent-U1charge}), and $g_{\mu\tau}$ is the $U(1)_{L_\mu-L_\tau}$ gauge coupling. Finally, the kinetic term for the extra gauge boson $Z_{\mu\tau}$ is given by the last term in Eq. \eqref{eqn:lag} with its field strength tensor $F_{\mu \tau}^{\alpha \beta} = \partial^\alpha Z_{\mu\tau}^\beta-\partial^\beta Z_{\mu\tau}^\alpha$.

In general, the Lagrangian \eqref{eqn:lag} may include the gauge kinetic mixing term \cite{Holdom:1985ag},
\begin{align}
\mathcal{L} \supset \frac{\zeta}{2}F_{\mu\tau}^{\alpha\beta}F_{\alpha\beta}\,,
\end{align}
between the $U(1)_{L_\mu - L_\tau}$ gauge boson and the SM $U(1)_Y$ gauge boson whose field-strength tensor is denoted by $F_{\alpha\beta}$. In the presence of the gauge kinetic mixing term, one may work with the physical gauge boson states instead of the original gauge boson states by diagonalising the mass matrix of the gauge bosons \cite{Babu:1997st}. Furthermore, as we shall see shortly, the DM phenomenology as well as the FOPT-associated GWs are qualitatively indifferent to the gauge kinetic mixing term.
Therefore, since the kinetic mixing term does not play an important role in our discussion, we assume, for simplicity, that $\zeta \ll 1$ in this work.\footnote{
In Refs. \cite{Altmannshofer:2019zhy,Biswas:2021dan}, it was shown that small values of the kinetic mixing parameter $\zeta$ are favoured from the muon $g-2$ aspect when taking into account the experimental constraint of Borexino \cite{Harnik:2012ni,Borexino:2017rsf}. See also, \textit{e.g.}, Ref. \cite{Bauer:2018onh} for a comprehensive study on experimental constraints on the kinetic mixing parameter $\zeta$.
}

\begin{center}
\begin{table}[t!]
\begin{tabular}{||c|c|c|c||}
\hline
\hline
\begin{tabular}{c}
Gauge\\
Group\\ 
\hline
${\rm SU(2)}_{\rm L}$\\ 
\hline
${\rm U(1)}_{\rm Y}$\\ 
\end{tabular}
&
\begin{tabular}{c|c|c}
\multicolumn{3}{c}{Baryon Fields}\\ 
\hline
$Q_{L}^{i}=(u_{L}^{i},d_{L}^{i})^{T}$&$u_{R}^{i}$&$d_{R}^{i}$\\ 
\hline
$2$&$1$&$1$\\ 
\hline
$1/6$&$2/3$&$-1/3$\\ 
\end{tabular}
&
\begin{tabular}{c|c|c}
\multicolumn{3}{c}{Lepton Fields}\\
\hline
$L_{L}^{i}=(\nu_{L}^{i},e_{L}^{i})^{T}$ & $e_{R}^{i}$ & $N_{R}^{i}$\\
\hline
$2$&$1$&$1$\\
\hline
$-1/2$&$-1$&$0$\\
\end{tabular}
&
\begin{tabular}{c|c|c|c}
\multicolumn{4}{c}{Scalar Fields}\\
\hline
$\phi_{h}$&$\phi_{H}$&$\phi_{1}$&$\phi_{2}$\\
\hline
$2$&$1$&$1$&$1$\\
\hline
$1/2$&$0$&$0$&$0$\\
\end{tabular}\\
\hline
\hline
\end{tabular}
\caption{Particle contents and their corresponding
charges under the SM gauge group.}
\label{tab:ParticleContent-SMcharge}
\end{table}
\end{center}
\begin{center}
\begin{table}[t!]
\begin{tabular}{|c|c|c|c|}
\hline
\hline
\begin{tabular}{c}
Gauge\\
Group\\ 
\hline
$U(1)_{L_\mu-L_\tau}$\\ 
\end{tabular}
&
\begin{tabular}{c}
\multicolumn{1}{c}{Baryon Fields}\\ 
\hline
$(Q^{i}_{L}, u^{i}_{R}, d^{i}_{R})$\\ 
\hline
$0$ \\ 
\end{tabular}
&
\begin{tabular}{c|c|c}
\multicolumn{3}{c}{Lepton Fields}\\ 
\hline
$(L_{L}^{e}, e_{R}, N_{R}^{e})$ & $(L_{L}^{\mu}, \mu_{R},
N_{R}^{\mu})$ & $(L_{L}^{\tau}, \tau_{R}, N_{R}^{\tau})$\\ 
\hline
$0$ & $1$ & $-1$\\ 
\end{tabular}
&
\begin{tabular}{c|c|c|c}
\multicolumn{4}{c}{Scalar Fields}\\
\hline
$\phi_{h}$ & $\phi_{H}$ & $\phi_{1}$& $\phi_{2}$ \\
\hline
$0$ &$1$ & $3 n_{\mu \tau}$ & $n_{\mu \tau}$\\
\end{tabular}\\
\hline
\hline
\end{tabular}
\caption{Particle contents and their corresponding
charges under $U(1)_{L_{\mu}-L_ {\tau}}$.}
\label{tab:ParticleContent-U1charge}
\end{table}
\end{center}

\vspace{-2cm}

The presence of the interaction term between $\phi_h$ and $\phi_H$ in Eq. \eqref{eqn:lagINT} introduces a mass mixing. 
In unitary gauge, the Higgs fields $\phi_{h}$ and $\phi_{H}$ after the spontaneous breaking of $SU(2)_L \times U(1)_Y \times U(1)_{L_\mu - L_\tau}$ gauge symmetry may be expressed as
\begin{align}
\phi_{h}=
\begin{pmatrix}
0 \\
\frac{v+H}{\sqrt{2}}
\end{pmatrix}
\qquad \text{and} \qquad
\phi_{H}=
\begin{pmatrix}
\frac{v_{\mu\tau} + H_{\mu\tau}}{\sqrt{2}}
\end{pmatrix}\,,
\label{eqn:phih}
\end{align}
where $v$ and $v_{\mu\tau}$ are the VEVs of the Higgs fields $\phi_h$ and $\phi_H$, respectively. 
The scalar mass matrix is then given by
\begin{align}
\mathcal{M}^2_{\rm scalar} = \left(\begin{array}{cc}
2\lambda_h v^2 & \lambda_{hH} v_{\mu\tau} v \\
\lambda_{hH} v_{\mu\tau} v & 2 \lambda_H v_{\mu\tau}^2
\end{array}\right) \,.
\label{eqn:mass-matrix}
\end{align}
In the presence of the Higgs-portal coupling $\lambda_{hH}$, the physical states are obtained after diagonalising the matrix $\mathcal{M}^2_{\rm scalar}$.
The mass eigenstates $h_1$ and $h_2$ can be written as
\begin{align}
h_{1} = H \cos \theta + H_{\mu\tau} \sin \theta\,, 
\quad
h_{2} = - H \sin \theta + H_{\mu\tau} \cos \theta\,.
\end{align}
The mixing angle $\theta$ and the mass eigenvalues $M^2_{h_1}$ and $M^2_{h_2}$ are given by
\begin{align}
\tan 2\theta &= \frac{\lambda_{hH}v_{\mu\tau} v}
{\lambda_h v^2 - \lambda_H v_{\mu\tau}^2}\,,
\label{eqn:mixingangle}
\\
M^2_{h_1} &= \lambda_h v^2 + \lambda_H v_{\mu\tau}^2 - 
\sqrt{(\lambda_h v^2 - \lambda_H v_{\mu\tau}^2)^2 
+ (\lambda_{hH}vv_{\mu\tau})^2}
\,,\label{eqn:massh1}
\\
M^2_{h_2} &= \lambda_h v^2 + \lambda_H v_{\mu\tau}^2 +
\sqrt{(\lambda_h v^2 - \lambda_H v_{\mu\tau}^2)^2 
+ (\lambda_{hH}vv_{\mu\tau})^2}
\,.
\label{eqn:massh2}
\end{align}
We identify the lighter scalar field $h_1$ with the observed SM Higgs field.

For the masses of the WIMP and FIMP, we obtain, with $\mu^2_{1,2} > 0$, as
\begin{align}
M^2_{1} = 
\mu^2_1 + \lambda_{1h} \frac{v^{2}}{2} 
+ \lambda_{1H} \frac{v^{2}_{\mu\tau}}{2} 
\,,\quad
M^2_{2} = 
\mu^2_2 + \lambda_{2h} \frac{v^{2}}{2} 
+ \lambda_{2H} \frac{v^{2}_{\mu\tau}}{2}
\,.
\end{align}

We summarise the particle contents of our model and their corresponding charges in Table \ref{tab:ParticleContent-SMcharge} and Table \ref{tab:ParticleContent-U1charge}.
In the remaining part of this section, we present the standard type-I seesaw mechanism that we adopt to explain the neutrino masses, and we briefly explain how the muon $g-2$ tension can be relieved in our model. For a detailed explanation, readers may refer to {\it e.g.} Refs. \cite{Biswas:2016yan,Biswas:2016yjr}.

%%%%%%%%%%%%%%%%
\subsection{Neutrino masses}
%%%%%%%%%%%%%%%%

Once the SM and $U(1)_{L_\mu - L_\tau}$ Higgs fields develop VEVs, the RH neutrino mass matrix can be expressed as
\begin{align}
M_{R} = 
\begin{pmatrix}
M_{ee} & \frac{h_{e\mu} v_{\mu\tau}}{\sqrt{2}} & \frac{h_{e\tau} v_{\mu\tau}}{\sqrt{2}}\\
\frac{h_{e\mu} v_{\mu\tau}}{\sqrt{2}} & 0 & M_{\mu\tau} e^{i\eta} \\
\frac{h_{e\tau} v_{\mu\tau}}{\sqrt{2}} & M_{\mu\tau} e^{i\eta} & 0
\end{pmatrix}
\,,
\end{align}
where $\eta$ is the only fermionic phase factor that cannot be absorbed by field redefinitions, and we see from the Yukawa terms in Eq. \eqref{eqn:lagN} that the Dirac mass matrix can be written as
\begin{align}
M_{D} =
\begin{pmatrix}
\frac{y_{e} v}{\sqrt{2}} & 0 & 0\\
0 & \frac{y_{\mu} v}{\sqrt{2}} & 0\\
0 & 0 & \frac{y_{\tau} v}{\sqrt{2}}
\end{pmatrix}
\,.
\end{align}
Therefore, the complete neutrino mass matrix is a $6 \times 6$ matrix in the basis $(\nu_{l}, N_{l})$,
\begin{align}
M_{\nu} =
\begin{pmatrix}
0 & M_{D}\\
M^{T}_{D} & M_{R}
\end{pmatrix}
\,,
\end{align}
After diagonalisation, we can obtain the mass matrix for the mass eigenstates. Then, we can write the light neutrino mass and heavy mass matrix as follows:
\begin{align}
m^{\rm light}_{\nu} = - M^{T}_{D} M^{-1}_{R} M_{D} 
\,, \quad
M^{\rm heavy}_{R} = M_{R}\,.
\end{align}
With the RH neutrino mass matrix elements in GeV range and the Dirac mass matrix in keV range, one may easily obtain the neutrino mass in the correct experimental range \cite{Esteban:2020cvm}; see also Refs. \cite{Biswas:2016yan,Biswas:2016yjr} for details. 

The RH neutrino mass matrix squared, $(M_R^{\rm heavy})^2$, can be diagonalised analytically when $M_{ee} = M_{\mu\tau}$ and $h_{e\tau} = h_{e\mu}$, and we obtain the eigenvalues as
\begin{align}
(M_{ee} - h_{e\tau} v_{\mu\tau})^2 \,,\quad
M_{ee}^2 \,,\quad
(M_{ee} + h_{e\tau} v_{\mu\tau})^2 \,.
\end{align}
In the following, we assume that this is the case.

%%%%%%%%%%%%%%%%
\subsection{Muon $g-2$}
%%%%%%%%%%%%%%%%

%%
\begin{figure}[t!]
\centering
\includegraphics[angle=0,height=8cm,width=8cm]{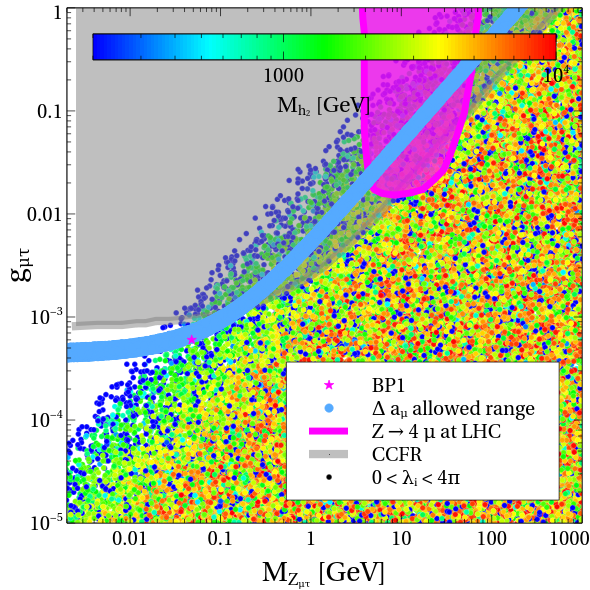}
\caption{Constraints on $g_{\mu\tau}$ and $M_{Z_{\mu\tau}}$. The cyan region relieves the muon $g-2$ tension. The magenta region is excluded by the $Z\rightarrow 4\mu$ searches from LHC \cite{CMS:2012bw,ATLAS:2014jlg,CMS:2018yxg}, while the grey region is excluded by the neutrino trident experiments CHARM-II and CCFR \cite{CHARM-II:1990dvf,CCFR:1991lpl,Altmannshofer:2014pba}. See also Ref. \cite{Chun:2018ibr} for constraints coming from the lepton universality test and LEP searches. The colour of the scan points represents $M_{h_2}$. The star ($*$) corresponds to our benchmark point 1 (see Table \ref{tab:GWbenchamarkpoints} in Section \ref{sec:FOPTGW}).}
\label{fig:muong2}
\end{figure}

The presence of additional gauge boson $Z_{\mu\tau}$ can alleviate the $(g-2)_{\mu}$ anomaly through the one-loop contribution, resulting in \cite{Gninenko:2001hx,Baek:2001kca}
\begin{align}
\Delta a_{\mu} = \frac{g^2_{\mu\tau}}{8 \pi^{2}} \int^{1}_{0} \frac{2 x (1 - x)^{2}}{(1-x)^{2} + r x} d x \,,
\end{align}
where $r = M^2_{Z_{\mu\tau}}/m^2_{\mu}$\,.
Figure \ref{fig:muong2} shows the region that addresses the discrepancy between the experimental and theoretical values of muon $g-2$, together with constraints from the neutrino trident experiments such as CHARM-II \cite{CHARM-II:1990dvf} and  CCFR \cite{CCFR:1991lpl,Altmannshofer:2014pba} and the LHC $Z\rightarrow 4\mu$ searches \cite{CMS:2012bw,ATLAS:2014jlg,CMS:2018yxg}. We observe that $M_{Z_{\mu\tau}} \lesssim 0.1$ GeV region with $4 \times 10^{-4} \lesssim g_{\mu\tau} \lesssim 1 \times 10^{-3}$ successfully explains the muon $g-2$ tension. We also present our scan points whose colour represents the value of $M_{h_2}$. One may clearly see from Fig. \ref{fig:muong2} that $M_{h_2} \gtrsim 1.1$ TeV is disfavoured from the muon $g-2$ point of view as long as the quartic couplings are in the perturbative regime. Therefore, throughout the paper, we consider $M_{h_2} \lesssim 1.1$ TeV. The star ($*$) in Fig. \ref{fig:muong2} depicts our benchmark point 1 (see Table \ref{tab:GWbenchamarkpoints}). Strong GW signals can be emitted from a region that explains the muon $g-2$ tension. In Section \ref{sec:FOPTGW}, we discuss possible GW signals in detail.

%%%%%%%%%%%%%%%%%%%%%%%%%%%%%%%%%%%
\section{Two-Component Dark Matter}
\label{sec:TCDM}
%%%%%%%%%%%%%%%%%%%%%%%%%%%%%%%%%%%

In this section, we examine the possibility of having DM component(s) in the present model. As the model contains two scalar DM candidates, we may have a single-component or two-component DM scenario depending on the mass range of the WIMP and FIMP DM particles.
We first discuss the production of DM when the $\mu$ term in Eq. \eqref{eqn:lagDM} is dominant and the other quartic terms associated with
the DM are also significant. 
We also look at the scenario when the $\mu$ parameter is less significant and quartic terms are the ones which take part in the DM productions.
In regime I and regime II, we study the effect of the $\mu$ term on the production of DM. In these cases, depending on the mass range of the WIMP DM, we obtain both the single-component and two-component DM scenarios.
In regime III, we study DM productions when the $\mu$ term is small and the quartic terms are relevant. In this regime, we have a two-component 
DM scenario where one component is WIMP-type DM and another component is FIMP-type DM.
In studying DM phenomenology, we have implemented our model in \textsc{FeynRules} \cite{Alloul:2013bka} and generated the \textsc{CalcHEP} files \cite{Belyaev:2012qa}. We have then used \textsc{micrOMEGAs} \cite{Belanger:2018ccd} to solve the coupled Boltzmann equations relevant for our study. Some useful analytical expressions are derived and summarised in Appendix \ref{apdx:analyticFI}.
We discuss the different regimes in detail below.

%%%%%%%%%%%%%%%%
\subsection{Regime I ($M_{h_2} < 2 M_{2}$ and $M_{1} > 3 M_{2}$)}
%%%%%%%%%%%%%%%%
As the WIMP DM mass is larger than three times the mass of the FIMP DM, a three-body decay channel from the WIMP DM to the FIMP DM is open in this regime. Since the WIMP DM decays into the FIMP DM, this regime gives us a single-component DM scenario, unless the lifetime is larger than the age of the Universe. It would require extremely small couplings to make the lifetime larger than the age of the Universe, and we do not consider such a scenario.
Additionally, we assume that the SM and BSM Higgs masses are such that the decay production of the FIMP DM is kinematically forbidden. Nevertheless, a freeze-in contribution through annihilation processes, $A B \rightarrow \phi^{\dagger}_{2} \phi_{2}$, where $A$ and $B$ are the SM particles, will be there. 

The Boltzmann equations associated with the WIMP and FIMP DM productions are given by
\begin{align}
\frac{d Y_{1}}{d x} &= 
-\frac{2 \pi^2}{45} 
\frac{M_{\rm Pl} M_{1} \sqrt{g_{*}(x)}}{1.66 x^{2}}
\langle \sigma v \rangle_{\rm th}
\left( Y^{2}_{1} - Y^{{\rm eq}2}_{1} \right) 
- \frac{3 M_{\rm Pl} x \sqrt{g_{*}(x)} }{1.66 M^{2}_{1} g_{s}(x)} 
\langle \Gamma \rangle \left( Y_{1} - Y^{3}_{2} \right) 
\,, \nonumber \\ 
\frac{d Y_{2}}{d x} &= 
\frac{4 \pi^2}{45} 
\frac{M_{\rm Pl} M_{1} \sqrt{g_{*}}}{1.66 x^{2}}
\sum_{i,j \in {\rm SM}, \phi_1 } 
\langle \sigma v \rangle_{ij} 
\left( Y^{{\rm eq}}_{i} Y^{{\rm eq}}_{j}  - Y^{2}_{2} \right) 
+ \frac{3 M_{{\rm Pl}} x \sqrt{g_{*}(x)}}{1.66 M^{2}_{1} g_{s}(x)} 
\langle \Gamma \rangle \left( Y_{1} - Y^{3}_{2} \right) 
\,,
\label{eqn:BE-single-component}
\end{align}
where $M_{\rm Pl} = 1.22 \times 10^{19}$ GeV is the Planck mass, and $g_{*}(x)$ and $g_{s}(x)$ are the effective and entropic degrees of freedom of the Universe.
Here, $Y_{1,2}\equiv n_{\phi_{1,2}}/S$ are the yields, with $n_{\phi_{1,2}}$ being the number densities and $S$ the entropy density.
The first equation corresponds to the evolution of the WIMP DM and the second equation represents the production of the FIMP DM.
In the right hand side of the first equation, the first term is the annihilation of the WIMP DM to the SM particles. Here, $\langle \sigma v \rangle_{\rm th}$ is the thermal average of cross section times velocity of DM annihilating to the SM particles.
The second term implies the three-body decay of the WIMP DM to the FIMP DM, where $\langle \Gamma \rangle$ is the thermal average of the decay rate $\Gamma$, defined as $\langle \Gamma \rangle = \Gamma K_{1}(x)/K_{2}(x)$ with $K_{1,2}$ being the modified Bessel functions of the second kind. The analytical expression for the three-body decay is provided in Appendix \ref{apdx:analyticFI}; see Eq. \eqref{eqn:three-body-decay}.
Similarly, the first term in the right hand side of the second equation represents the annihilation contribution to the FIMP DM and the second term is the decay contribution of the WIMP DM to the FIMP DM. Here, $\langle \sigma v \rangle_{ij}$ is the thermal average of annihilations of $i,j$ particles to FIMP DM.
Due to the allowed decay term of the WIMP to the FIMP, we see that the WIMP DM eventually decays to the FIMP DM before big bang nucleosynthesis (BBN). We still do not have any contribution to visible energy even when the WIMP DM decays after BBN, and thus, our model remains safe from the constraints which come from light elements abundances \cite{Kawasaki:2017bqm}.

\begin{figure}[t!]
\centering
\includegraphics[angle=0,height=7cm,width=7cm]{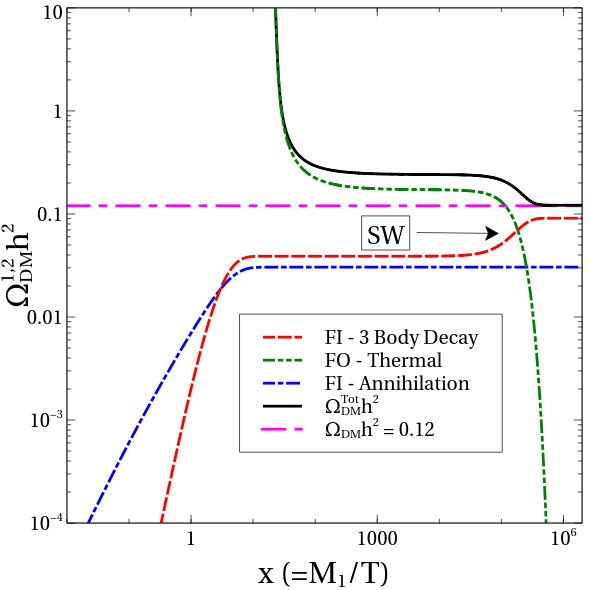}
\caption{Evolution of DM relic density produced by different mechanisms: freeze-in production from three-body decay (red dashed line), thermal freeze-out of WIMP (green double-dot-dashed line), freeze-in production from annihilation (blue dot-dashed line), and the total contribution (black solid line). The model parameters are chosen as follows: $M_{1} = 1650$ GeV, $M_{2} = 500$ GeV, $M_{h_2} = 500$ GeV, $M_{Z_{\mu\tau}} = 0.1$ GeV, $g_{\mu\tau} = 9 \times 10^{-4}$, $\lambda_{2i} = 6 \times 10^{-12}$ ($i = h, H$), $\lambda_{12} = 6 \times 10^{-12}$, $\lambda_{1i} = 0.5$ ($i = h, H$), and $\mu = 7.5 \times 10^{-11}$. The magenta dot-dashed line corresponds to the correct value of DM relic density.}
\label{fig:bf-different-contribution}
\end{figure}

In Fig. \ref{fig:bf-different-contribution}, we show the evolution of the DM relic density in terms of $x = M_1/T$. We examine the contributions of different production mechanisms. The blue dot-dashed line in Fig. \ref{fig:bf-different-contribution} corresponds to the production of the FIMP DM through annihilation processes which saturate at $x \simeq 1$. The dominating processes in the annihilation contribution are the four-point contact terms which are $A B \rightarrow \phi^{\dagger}_{2} \phi_2$ ($A, B = h_{1,2}, \phi_{1}$) and are not propagator-suppressed. The green double-dot-dashed line represents the evolution of the WIMP DM which freezes out at $x \simeq 20$ and starts to decay into the FIMP DM at $x \simeq 10^{5}$. Since the WIMP DM decays into the FIMP DM, this regime corresponds to a single-component DM scenario. The red dashed line is the freeze-in production from the three-body decay of the WIMP DM {\it i.e.} $\phi_{1} \rightarrow \phi_{2} \phi_{2} \phi_{2}$ that happens at $x \simeq 1$ which means that the WIMP DM is in thermal equilibrium with the cosmic soup.
Moreover, there is also another contribution that is superWIMP (SW) contribution appears at $x \simeq 10^{5}$ \cite{Feng:2003xh}. It comes from the three-body decay of the WIMP DM. Finally, the black solid line corresponds to the total DM relic density which comes after summing all the contributions. 
The magenta dot-dashed line corresponds to the correct value of the DM relic density. We see that, for the choice of model parameters outlined in the caption of Fig. \ref{fig:bf-different-contribution}, our model correctly produces the exact amount of DM relic density.    

\begin{figure}[t!]
\centering
\includegraphics[angle=0,height=7cm,width=7cm]{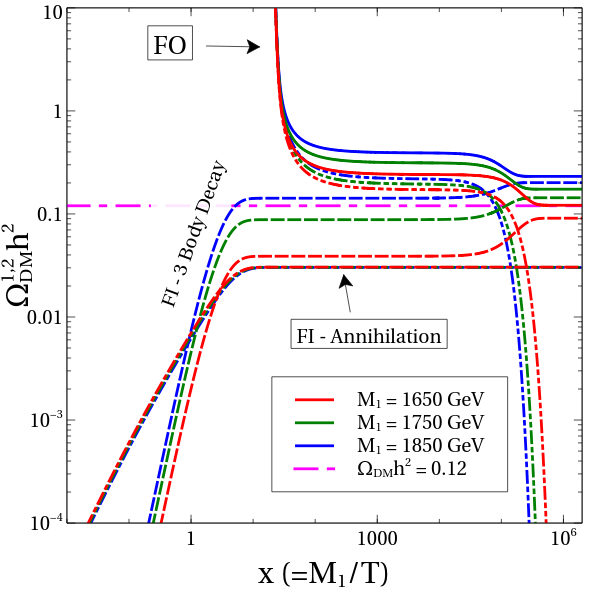}
\qquad
\includegraphics[angle=0,height=7cm,width=7cm]{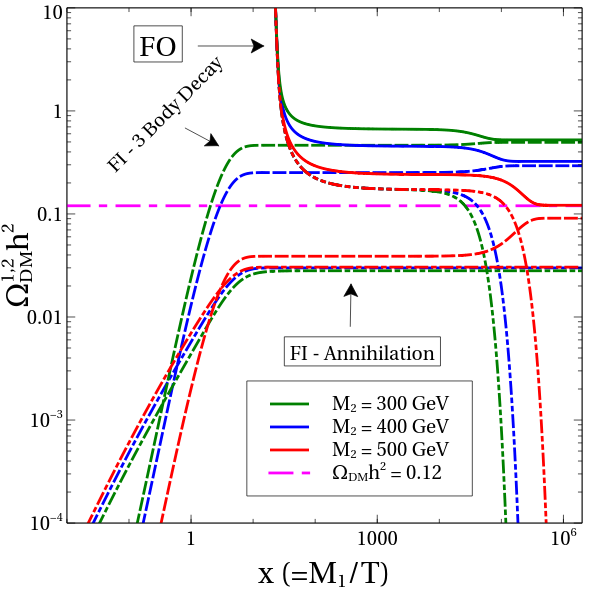}
\caption{Evolutions of DM relic density for three different values of the WIMP DM mass $M_1$ ({\it left}) and for three different values of the FIMP DM mass $M_2$ ({\it right}). 
For the rest of the model parameters, see Fig. \ref{fig:bf-different-contribution}.}
\label{fig:regime-I-vary-mfimp-mwimp}
\end{figure}

Changes in the DM relic density with respect to the masses of the WIMP DM and FIMP DM are shown respectively in the left panel and the right panel of Fig. \ref{fig:regime-I-vary-mfimp-mwimp}.
The freeze-in production of the FIMP DM due to the annihilation is insensitive to the WIMP DM mass. This is consistent with the observation that there is no direct effect of the WIMP DM mass on the annihilation production of the FIMP DM apart from the annihilation process $\phi^{\dagger}_{1} \phi_{1} \rightarrow \phi^{\dagger}_{2} \phi_{2}$ which has negligible dependence on the WIMP DM mass. 
The freeze-in production from the decay of the WIMP DM when it is in thermal equilibrium, {\it i.e.}, $x \lesssim 20$, is in general inversely proportional to the WIMP DM mass in this regime. However, we observe the opposite behaviour, {\it i.e.}, we get more production as the WIMP DM mass increases. This is due to the fact that, for a low value of the WIMP DM mass, $M_{1} = 1650$ GeV, we have a phase-space suppression in the decay. Thus, we get less amount of the FIMP DM, and when we increase the WIMP DM mass, the effect of phase space gets reduced, and we obtain more DM from decay. 
The WIMP DM freezes out at $x \simeq 20$ and starts to decay into the FIMP DM at $x \simeq 10^5$. We see that the WIMP DM starts to decay into the FIMP DM earlier as the WIMP DM mass increases as the double-dot-dashed lines in Fig. \ref{fig:regime-I-vary-mfimp-mwimp} indicate. This is because the decay width is linearly proportional to the WIMP DM mass. On top of that, there is also the phase-space suppression which further reduces the decay width and delays the WIMP decay. 
When the decay of the WIMP DM happens, we see a rise in the production of the FIMP DM at $x \simeq 10^{5}$ which is similar to the superWIMP production mechanism.

In the right panel of Fig. \ref{fig:regime-I-vary-mfimp-mwimp}, we show the dependence on the FIMP DM mass. For the FIMP production due to the annihilation which is represented by the dot-dashed lines, we see a slight variation in the relic density. This is because $h_{2} h_{2} \rightarrow \phi^{\dagger}_2 \phi_2$ is the dominant process, and we have taken the BSM Higgs mass to be $M_{h_2} = 500$ GeV which is comparable to the FIMP DM mass. Therefore, suppression due to the phase-space factor and increment due to mass compensate each other.
The FIMP DM production due to the three-body decay is shown by the dashed lines. We observe one order of magnitude difference in the DM production when we vary the FIMP DM mass from 500 GeV to 400 GeV. This happens purely because the effect of phase space is small, and the same effect continues when we decrease the FIMP DM mass further. 
Finally, let us discuss the production of the WIMP DM which decays into the FIMP DM at $x \simeq 10^{5}$. Again, we see that, as the FIMP DM mass decreases from 500 GeV to 400 GeV, WIMP decay width increases due to lower phase-space suppression which indicates that the WIMP decays earlier. This is visible by the double-dot-dashed lines. When the WIMP DM decay happens, we have further production of the FIMP DM similar to the superWIMP production.
In both the left and right panels of Fig. \ref{fig:regime-I-vary-mfimp-mwimp}, solid lines represent the total DM relic density after summing all the production contributions.

\begin{figure}[t!]
\centering
\includegraphics[angle=0,height=7cm,width=7cm]{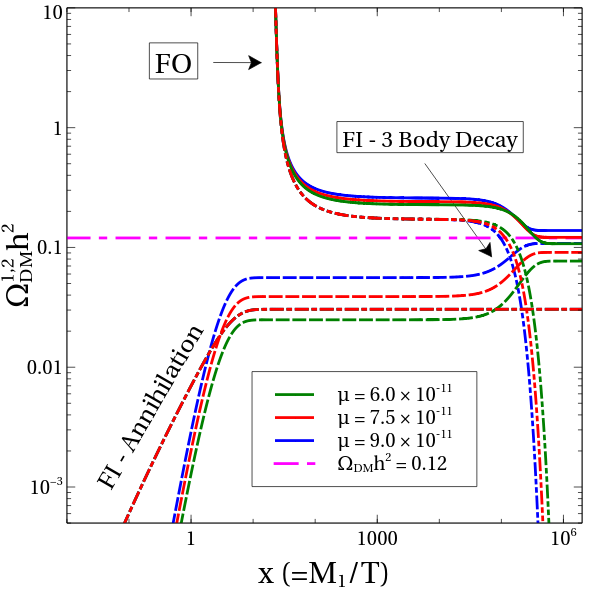}
\qquad
\includegraphics[angle=0,height=7cm,width=7cm]{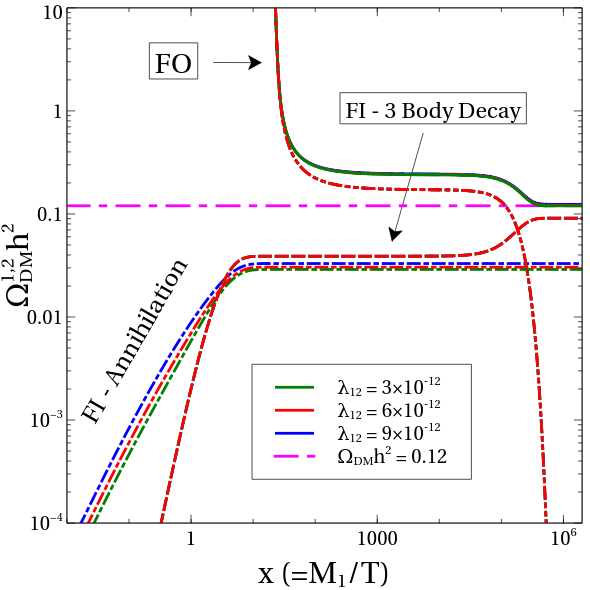}
\caption{Dependence of the DM relic density on the $\mu$ ({\it left}) and $\lambda_{12}$ ({\it right}) parameters. For the rest of the model parameters, see Fig. \ref{fig:bf-different-contribution}.}
\label{fig:regime-I-vary-mu-lamdm2dm1}
\end{figure}

The left and right panels of Fig. \ref{fig:regime-I-vary-mu-lamdm2dm1}, shows the dependence of DM relic density for three different values of the $\mu$ and $\lambda_{12}$ parameters, respectively. 
In the left panel, from the freeze-in contribution through annihilation, we see that there is no change in the relic density coming from the annihilation contribution which is represented by the dot-dashed line. This is because the annihilation process associated with the $\mu$ term is proportional to $Y^{{\rm eq}}_{1} Y_{2}$ ($Y_{2}= 0$ at initial value of $x$), while other annihilation terms are proportional to $Y^{{\rm eq}}_{A} Y^{{\rm eq}}_{B}$ ($A, B$ are the annihilating particles). 
Let us turn to the production of the FIMP DM from the three-body decay of the WIMP DM. The production of the FIMP DM before $x \simeq 10$ occurs in the domain when the WIMP DM is still in thermal equilibrium, and from its decay, the FIMP DM is produced. We clearly see that the production has a quadratic dependence on the $\mu$ parameter which is perfectly consistent with the analytical expression given in the Appendix \ref{apdx:analyticFI}; see Eq. \eqref{eqn:three-body-decay}. 
We note that the freeze-out temperature of the WIMP DM does not depend on the $\mu$ parameter while the WIMP DM decay does.
Decay of the WIMP happens earlier (later) for a higher (lower) value of $\mu$. This is consistent with the analytical expression; see Appendix \ref{apdx:analyticFI} for details.
Depending on the decay occurrence, the superWIMP contribution to the FIMP DM happens earlier or later and has an equal contribution for all three values, as the freeze-out contributions do not depend on the $\mu$ parameter, and this contribution is equal to $\Omega^{2}_{\rm DM}h^{2} = \Omega^{1}_{\rm DM}h^{2}(M_{2}/M_{1})$.

From the right panel of Fig. \ref{fig:regime-I-vary-mu-lamdm2dm1}, one may clearly see that the three-body decay and the superWIMP production do not depend on $\lambda_{12}$. However, we see changes in the production coming from the annihilation process, $\phi^{\dagger}_{1} \phi_{1} \rightarrow \phi^{\dagger}_{2} \phi_2$. 
In both the left and right panels of Fig. \ref{fig:regime-I-vary-mu-lamdm2dm1}, the solid lines correspond to the total contribution in DM relic density. 

\begin{figure}[t!]
\centering
\includegraphics[angle=0,height=7cm,width=7cm]{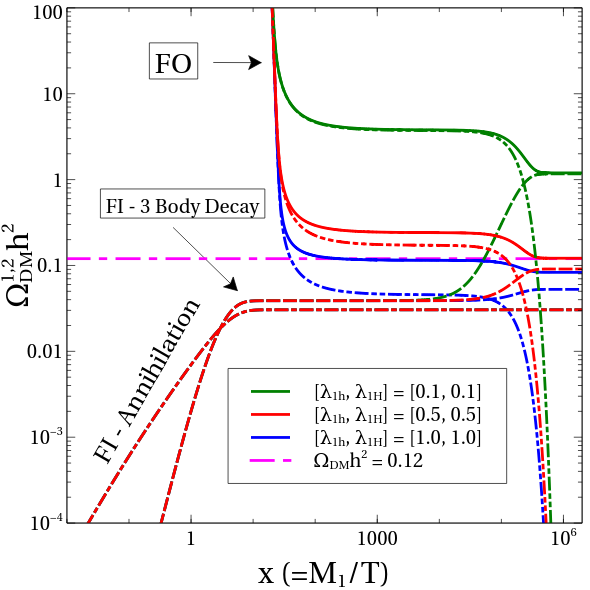}
\qquad
\includegraphics[angle=0,height=7cm,width=7cm]{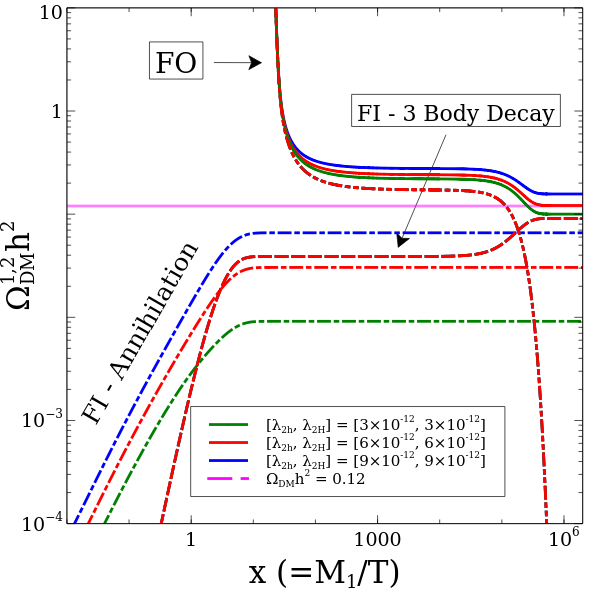}
\caption{Dependence of the DM relic density on the quartic coupling of the WIMP with the Higgses $\lambda_{1i}$ ($i = h, H$) ({\it left}) and the quartic coupling of the FIMP with Higgses $\lambda_{2i}$ ($i = h, H$) ({\it right}). For the rest of the model parameters, see Fig. \ref{fig:bf-different-contribution}.}
\label{fig:regime-I-vary-dm1-dm2-hi}
\end{figure}

Finally, the dependence of the DM relic density on the quartic couplings of the DM with the Higgses, $\lambda_{ij}$ ($i=1,2$ and $j=h,H$), is shown in Fig. \ref{fig:regime-I-vary-dm1-dm2-hi}.
From the left panel, we see that the production of the FIMP DM from the freeze-in by the three-body decay and annihilation does not change; see the red dashed and dot-dashed lines. We can also see that there is a significant change in the final value when we consider the freeze-out production of the WIMP DM. This can be explained in a very simple way. The WIMP relic density is determined from the inverse of thermal average of cross section times velocity, namely $\Omega^{1}_{\rm DM}h^{2} \propto 1/\lambda^2_{1i}$ ($i=h,H$).
Therefore, larger values of $\lambda_{1j}$ ($j=h,H$) imply that we face the situation when most of the particles annihilate away, and thus, we have less abundance for WIMP. 
The relative strength of the WIMP DM relic density due to different values of quartic couplings is given by $\Omega^{1}_{\rm DM} h^{2} |_{A}/\Omega^{1}_{\rm DM} h^{2}|_{B} = \lambda^2_{1i}|_{B}/\lambda^2_{1i}|_{A}$ ($i = h, H$). This is consistent with our numerical results as one may see from the blue, red, and green lines for the freeze-out production of the WIMP DM.
Since there is no variation of the FIMP DM production from the three-body decay and annihilation, most of the changes in the FIMP DM production comes after $x \simeq 10^{5}$ when the WIMP DM decays into the FIMP DM. Solid lines correspond to the total contribution after taking into
account all the production mechanisms.

On the other hand, from the right panel of Fig. \ref{fig:regime-I-vary-dm1-dm2-hi}, we see that the FIMP DM production from the three-body decay of the WIMP when the WIMP is in thermal equilibrium does not change as the quartic couplings vary. We can also see that the freeze-out production of the WIMP DM shown by the double-dot-dashed line is not affected by the change of the quartic couplings. However, the FIMP production from annihilation changes. This is indeed consistent with the observation that the freeze-in contribution by annihilation processes is proportional to the quartic couplings. In this case, the production is directly proportional to the quartic coupling which is visible by the dot-dashed lines; see Eqs. \eqref{eqn:scattY2-1} and \eqref{eqn:scattY2-2} in Appendix \ref{apdx:analyticFI}.

%%%%%%%%%%%%%%%%
\subsection{Regime II ($M_{h_2} < 2 M_{2}$ and $M_{1} < 3 M_{2}$)}
%%%%%%%%%%%%%%%%
When the WIMP DM mass is less than the three times the FIMP DM mass, the three-body decay channel of the WIMP DM to the FIMP DM is kinematically forbidden. Therefore, in this case, we have a two-component DM scenario, one WIMP-type DM and one FIMP-type DM. 

The WIMP DM freezes out at $x \simeq 20$, and we obtain relic density in the experimentally allowed range put by Planck \cite{Planck:2018vyg} near the Higgs resonance region. 
For the FIMP DM, we examine the effect of the $\mu$ term on the FIMP DM production and choose the FIMP DM mass in such a way that the decay channel $h_{2} \rightarrow \phi^{\dagger}_{2} \phi_2$ is kinematically forbidden. Nevertheless, we have an annihilation contribution in the production of the FIMP DM through the freeze-in mechanism. 
Due to the presence of the $\mu$ term, there exists $\phi_1 \phi_2 \rightarrow \phi_2 \phi_2$ annihilation process, and it will exhibits an exponential growth. In the production of the FIMP, at $x \simeq 0.01$, we have a tiny amount of FIMP DM produced from the annihilation processes of the SM particles, and at $x \simeq 1$, the exponential enhancement will take place which will be discussed in detail below. 

The governing Boltzmann equations in this regime are given by
\begin{align}
\frac{d Y_{1}}{d x} &= 
-\frac{2 \pi^2}{45} 
\frac{M_{{\rm Pl}} M_{1}}{1.66 x^{2}}
\sqrt{g_{*}} \langle \sigma v \rangle_{\rm th}
\left( Y^{2}_{1} - Y^{{\rm eq}2}_{1} \right)  
\,,\nonumber \\ 
\frac{d Y_{2}}{d x} &= 
-\frac{2 \pi^2}{45} \frac{M_{{\rm Pl}} M_{1}}{1.66 x^{2}}
\sqrt{g_{*}} \langle \sigma v \rangle_{\rm exp}
\left( Y^{2}_{2} - Y^{{\rm eq}}_{1} Y_{2} \right) 
+ \frac{4 \pi^2}{45} \frac{M_{{\rm Pl}} M_{1}}{1.66 x^{2}}
\sqrt{g_{*}} \sum_{i,j \in {\rm SM}, \phi_1} 
\langle \sigma v \rangle_{ij} 
\left( Y^{{\rm eq}}_{i} Y^{{\rm eq}}_{j} - Y^{2}_{2}\right) 
\,.
\label{fig:regime-II-BE}
\end{align}
The first (second) equation represents the evolution of the WIMP (FIMP) DM. Here, $\langle \sigma v \rangle_{\rm exp}$ is the thermal average associated with the exponential growth computed using the prescription described in Appendix \ref{apdx:exp-yield}. We now discuss the effect of model parameters on the production of WIMP and FIMP DM by different mechanisms.

\begin{figure}[t!]
\centering
\includegraphics[angle=0,height=7cm,width=7cm]{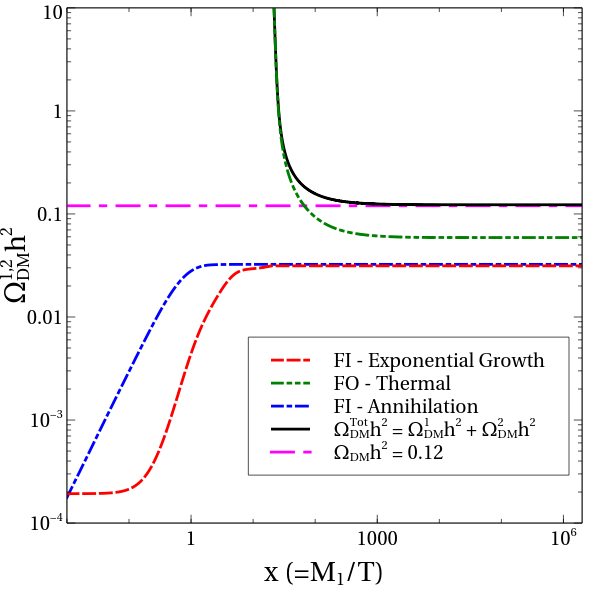}
\caption{Evolution of the DM relic density in regime II. In this regime, we have a two-component DM scenario with one WIMP DM and one FIMP DM. The evolution of the WIMP DM is shown in the green double-dot-dashed line, while the freeze-in productions of FIMP DM through annihilation and exponential growth are shown in the blue dot-dashed line and red dashed line, respectively. The black solid line corresponds to the sum of the WIMP and FIMP DM relic densities, and the magenta dot-dashed line indicates the observed value of DM relic density. 
The model parameters are chosen as $M_{1} = 500$ GeV, $M_{2} = 700$ GeV, $M_{h_2} = 1000$ GeV, $\lambda_{2i} = 6 \times 10^{-12}$ ($i = h,H$), $\lambda_{12} = 6 \times 10^{-12}$, $\lambda_{1h} = 0.05$, $\lambda_{1H} = 0.125$, and $\mu = 10^{-6}$. 
}
\label{fig:regime-II-bf-different-contribution}
\end{figure}

In Fig. \ref{fig:regime-II-bf-different-contribution}, the evolution of the WIMP and FIMP DM relic densities is shown. The model parameters are chosen in such a way that the WIMP and FIMP DM relic densities contribute equally and generate a total DM relic density in the correct ballpark value as referred by the Planck collaboration \cite{Planck:2018vyg}. 
The red dashed line in Fig. \ref{fig:regime-II-bf-different-contribution} corresponds to the exponential growth of FIMP DM due to the presence of the process $\phi_{1} \phi_{2} \rightarrow \phi_{2} \phi_2$. This kind of process can be solved analytically. The co-moving number density can be expressed as
\begin{align}
Y_{2} = Y^{\rm ini}_{2} 
e^{\int_{x_{\rm ini}}^{\infty} 
\frac{2 \pi^2}{45} \frac{M_{{\rm Pl}} M_{1}}{1.66 x^{2}}
\sqrt{g_{*}} \langle \sigma v \rangle Y^{{\rm eq}}_{1} d x}
\,,
\end{align}
where $Y^{\rm ini}_{2}=Y_2(x = x_{\rm ini})$. Thus, we see an exponential enhancement of the FIMP DM.
The blue dot-dashed line represents the FIMP DM production through the annihilation processes $A B \rightarrow \phi^{\dagger}_{2} \phi_2$ where
$A$ and $B$ are the particles in thermal equilibrium. The green double-dot-dashed line indicates the WIMP DM production through the freeze-out mechanism which happens at $x \simeq 20$. The total sum of the WIMP and FIMP DM contributions is depicted by the black solid line which matches with the correct value of DM relic density $\Omega_{\rm DM} h^{2} = 0.12$. 

\begin{figure}[t!]
\centering
\includegraphics[angle=0,height=7cm,width=7cm]{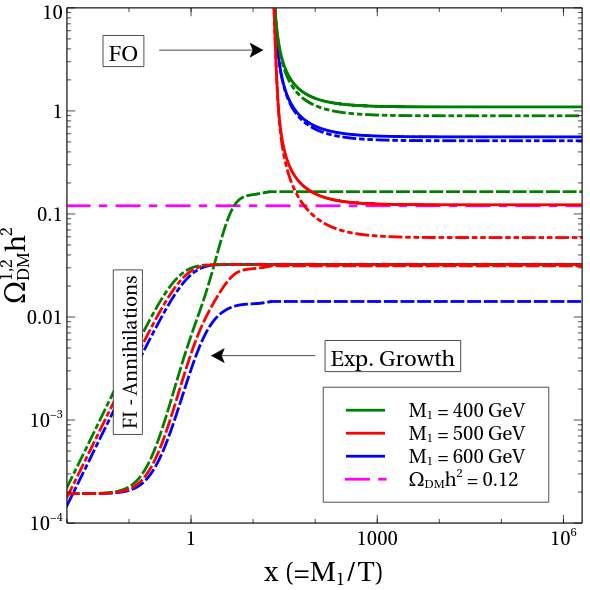}
\qquad
\includegraphics[angle=0,height=7cm,width=7cm]{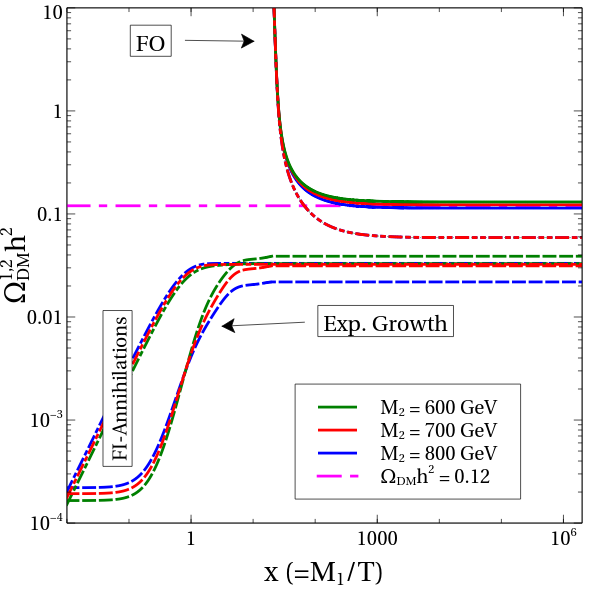}
\caption{Dependence of the DM relic density on the WIMP DM mass $M_1$ ({\it left}) and FIMP DM mass $M_2$ ({\it right}). For the rest of the model parameters, see Fig. \ref{fig:regime-II-bf-different-contribution}.
}
\label{fig:regime-II-vary-mfimp-mwimp}
\end{figure}

Figure \ref{fig:regime-II-vary-mfimp-mwimp} shows the dependence of DM relic density on the WIMP and FIMP DM masses. 
In the left panel, we see that the WIMP DM mass has no observable effect on the freeze-in production of FIMP DM through annihilation.
However, the WIMP DM mass affects the exponential growth of FIMP DM as the thermal average of cross section times velocity is inversely proportional to mass of the initial state particle, which is the WIMP DM in the present case.
In the case of the WIMP DM production, when we increase or decrease the WIMP DM mass around $ M_{1} = 500$ GeV, we get more abundance for the WIMP DM. This is understood from the resonance behaviour of the Higgs-mediated diagram. In Fig. \ref{fig:regime-II-vary-mfimp-mwimp}, we have considered $M_{h_2} = 1000$ GeV which is the resonance region for $M_{1} = 500$ GeV DM. Thus, we get a large annihilation cross section which results in the reduction in WIMP abundance. If the WIMP DM mass deviates from 500 GeV, we get a smaller value of annihilation cross section and higher WIMP DM abundance.
The solid lines correspond to the total contribution in DM relic density both from the WIMP and FIMP contributions.

In the right panel of Fig. \ref{fig:regime-II-vary-mfimp-mwimp}, we may observe the effect of FIMP DM mass on the production of WIMP and FIMP DM by different mechanisms.
The FIMP DM mass has little impact on the FIMP DM production through annihilation, while it has an observable effect on the exponentially enhanced production of FIMP DM. For $x \lesssim 1$, we see that the change in the FIMP DM relic density is proportional to the FIMP DM mass. On the other hand, for $x \gtrsim 1$, we get a similar kind of enhancement as discussed in the previous paragraph, due to the dependence of the thermal average of the cross section on mass.
As the chosen masses are in a large range, we see no big difference in the produced relic densities like before. There is also no effect of the FIMP DM mass on the production of WIMP DM through the freeze-out process. The solid lines are the total sum of WIMP and FIMP DM relic densities, and they all match the correct value of the DM relic density given by Planck. 

\begin{figure}[t!]
\centering
\includegraphics[angle=0,height=7cm,width=7cm]{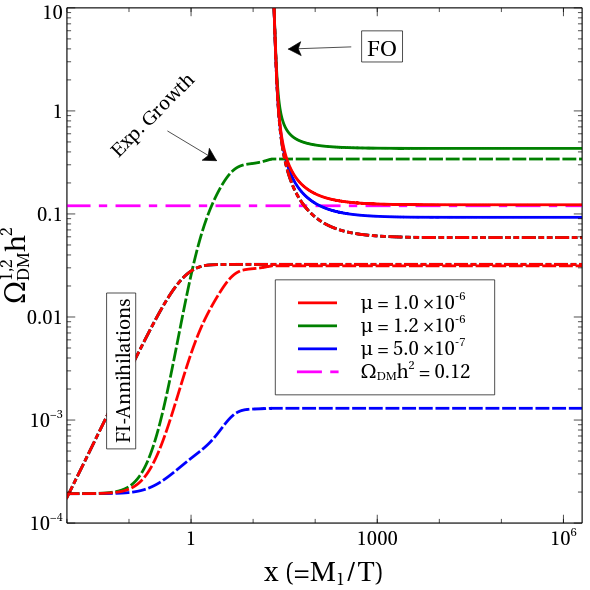}
\qquad
\includegraphics[angle=0,height=7cm,width=7cm]{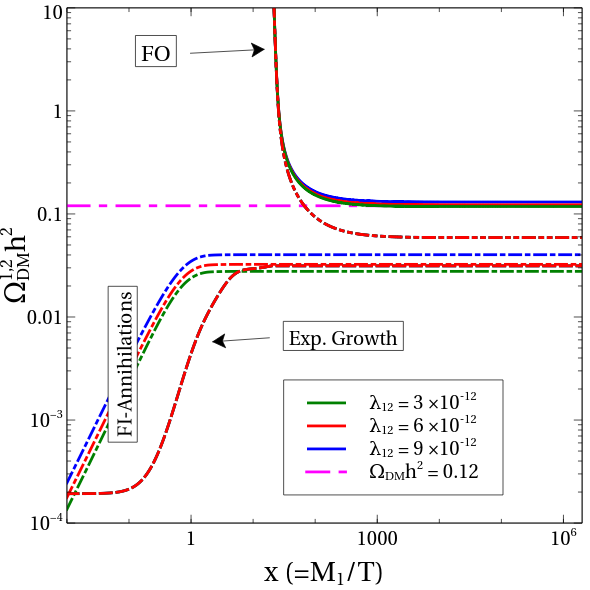}
\caption{Dependence of the DM relic density on the $\mu$ parameter ({\it left}) and the quartic coupling $\lambda_{12}$ between the WIMP DM and FIMP DM ({\it right}). For the rest of the model parameters, see Fig. \ref{fig:regime-II-bf-different-contribution}.
}
\label{fig:regime-II-vary-mu-lamdm2dm1}
\end{figure}

The dependences of the production of WIMP and FIMP DM on the $\mu$ parameter and the quartic coupling between the WIMP DM and FIMP DM $\lambda_{12}$ are respectively shown in the left and right panels of Fig. \ref{fig:regime-II-vary-mu-lamdm2dm1}.
The $\mu$ parameter only affects the process $\phi_{1} \phi_{2} \rightarrow \phi_{2} \phi_{2}$, and thus, other DM productions do not change.
Looking at the freeze-in production of FIMP DM from annihilation and WIMP DM production through the freeze-out process, we easily see that these production mechanisms do not vary when $\mu$ changes.
However, we can see a strong dependence of the exponential enhancement on the $\mu$ parameter. If we take $\mu \lesssim 10^{-7}$, then the exponential enhancement is absent, while for $\mu \gtrsim 5 \times 10^{-6}$, there exists a tremendous exponential enhancement in the production which overproduces the DM. Thus, higher values of $\mu$ are disfavoured. The solid lines are again the total sum of WIMP and FIMP contributions, and the variation in their values are solely due to the effect of the exponential enhancement. 

On the other hand, since the quartic coupling $\lambda_{12}$ is in the feeble regime, it does not contribute to the WIMP DM production which is clearly visible by the double-dot-dashed line which is same for all the three values of $\lambda_{12}$. The dashed line, which accounts for the FIMP DM production through the exponential enhancement, is also unchanged for different values of $\lambda_{12}$. However, the freeze-in contribution through annihilation depends on the $\lambda_{12}$ parameter as the dot-dashed lines indicate. The amount of DM production through annihilation depends quadratically on the $\lambda_{12}$ parameter.

\begin{figure}[t!]
\centering
\includegraphics[angle=0,height=7cm,width=7cm]{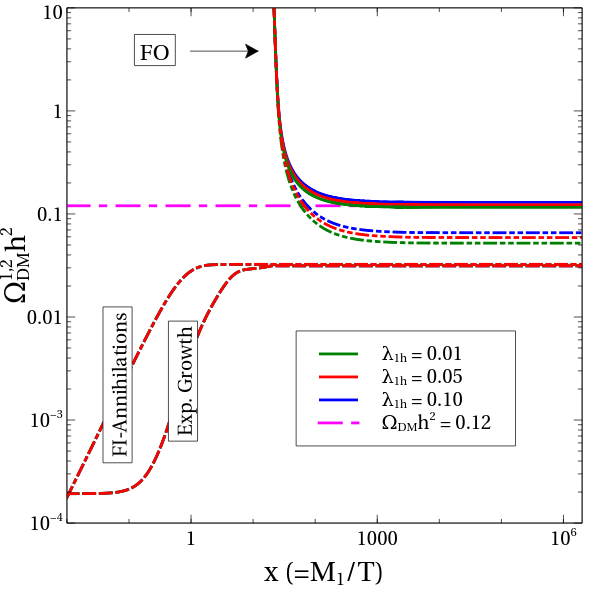}
\qquad
\includegraphics[angle=0,height=7cm,width=7cm]{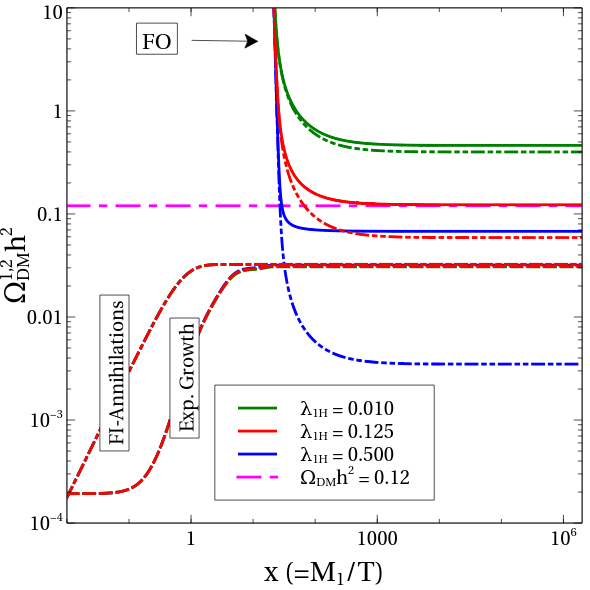}
\caption{
Dependence of the DM relic density on the quartic coupling between the WIMP DM and the SM Higgs ({\it left}) and the quartic coupling between the WIMP DM and the BSM Higgs ({\it right}).
For the rest of the model parameters, see Fig. \ref{fig:regime-II-bf-different-contribution}.
}
\label{fig:regime-II-vary-lam-dm1-h1}
\end{figure}

The left panel and the right panel of Fig. \ref{fig:regime-II-vary-lam-dm1-h1} show the dependence of the DM relic density on the quartic coupling between the WIMP DM and the SM Higgs $\lambda_{1h}$ and the quartic coupling between the WIMP DM and the BSM Higgs $\lambda_{1H}$, respectively.
The coupling $\lambda_{1h}$ connects the WIMP DM to the visible sector through the SM Higgs. Since this quartic coupling does not affect the exponential growth of the FIMP DM and has a negligible effect on the FIMP DM through annihilation, there is no change in the FIMP DM production for different values of $\lambda_{1h}$. We can see, however, changes in the WIMP DM production, although the difference is small. The small dependence on $\lambda_{1h}$ is due to the fact that the WIMP DM mass is chosen in such a way that it lies in the BSM Higgs resonance regime. Moreover, we have kept $\lambda_{1h}$ below $0.1$. Otherwise, the WIMP DM will be ruled out by the direct detection experiments. 

The right panel of Fig. \ref{fig:regime-II-vary-lam-dm1-h1} indicates that the quartic coupling $\lambda_{1H}$ has no effect on the FIMP DM production as well. However, we see that a change in $\lambda_{1H}$ results in an order of magnitude variation in the WIMP DM relic density. This is because our parameters are chosen such that the WIMP mass is in the resonance region for the second, BSM Higgs, $M_{1} \simeq M_{h_2}/2$. Therefore, a change in $\lambda_{1H}$ which measures the coupling strength for $\phi^{\dagger}_1 \phi_{1} h_{2}^2 $ has a direct impact on the thermal DM relic density.

\begin{figure}[t!]
\centering
\includegraphics[angle=0,height=7cm,width=7cm]{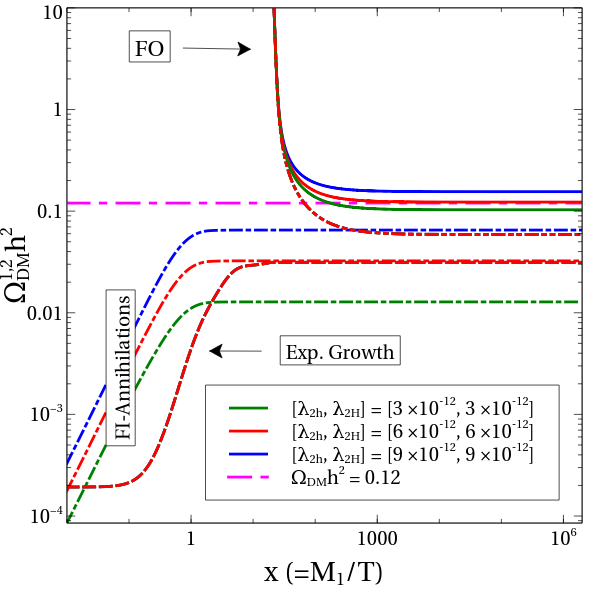}
\qquad
\includegraphics[angle=0,height=7cm,width=7cm]{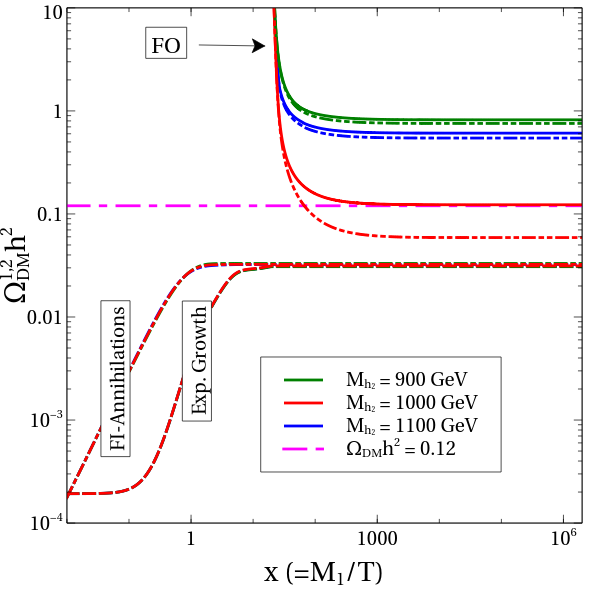}
\caption{
Dependence of the DM relic density on the quartic coupling between the FIMP DM and the SM and BSM Higgses ({\it left}) and the BSM Higgs mass ({\it right}).
For the rest of the model parameters, see Fig. \ref{fig:regime-II-bf-different-contribution}.
}
\label{fig:regime-II-vary-lambda-2-hH}
\end{figure}

The left panel of Fig. \ref{fig:regime-II-vary-lambda-2-hH} shows the dependence of the DM relic density on the quartic coupling between the FIMP DM and the SM and BSM Higgses $\lambda_{2i}$ ($i = h, H$).
Since $\lambda_{2h}$ and $\lambda_{2H}$ are associated with the FIMP DM, it does not affect the WIMP DM production as one may easily see from the figure.
The quartic coupling $\lambda_{2i}$ ($i = h, H$) also does not alter the FIMP DM production by the exponential enhancement. 
On the other hand, we see that the FIMP production by annihilation gets affected due to the variation of $\lambda_{2h}$ and $\lambda_{2H}$. This is because the associated annihilation processes $A B \rightarrow \phi^{\dagger}_2 \phi_2$, where $A$ and $B$ belong to the SM and BSM particles, directly depend on the strength of the $\lambda_{2h}$ and $\lambda_{2H}$ couplings. 
The changes in the solid line, which is the total sum of both the FIMP and WIMP contributions, are due to the variation in FIMP DM relic density coming from the annihilation part. 

The right panel of Fig. \ref{fig:regime-II-vary-lambda-2-hH} shows the dependence of the DM relic density on the BSM Higgs mass $M_{h_2}$.
Since in this regime, the decay process $h_2 \rightarrow \phi^{\dagger}_{2} \phi_2$ is not allowed, we do not see any observable effect on the FIMP DM production. However, we see an effect on the production of WIMP DM.
The reason is exactly the same as the one we discussed earlier for the left panel of Fig. \ref{fig:regime-II-vary-mfimp-mwimp}. Here as well, since $M_{1} = 500$ GeV, if $M_{h_2}$ deviates from $M_{h_2} = 1000$ GeV, we are basically going away from the resonance region. This means that DM freezes out earlier due to the reduction in the thermal cross section, and we get higher WIMP DM relic density. The changes in the solid line
are purely due to variation in the WIMP contribution to the DM relic density.

%%%%%%%%%%%%%%%%
\subsection{Regime III ($M_{h_2} > 2 M_{2}$ with $\mu$ negligible)}
%%%%%%%%%%%%%%%%
In this regime, one should take into account the FIMP DM production from the decay of the Higgses as well. Throughout the discussion, we assume that $\mu$ is negligible and focus on two-component DM scenarios.
Since $\mu$ is negligible, we may neglect the exponential enhancement in the FIMP DM production. We note that this scenario is different from the individual study of WIMP \cite{Rodejohann:2015lca,Biswas:2016ewm,Biswas:2016yan}  and FIMP \cite{Biswas:2016yjr} as the FIMP DM can also be produced from the annihilation of the WIMP DM through the process $\phi^{\dagger}_{1} \phi_1 \rightarrow \phi^{\dagger}_{2} \phi_2 $.
This annihilation contribution can be increased or decreased with the strength of the $\lambda_{12}$ parameter as discussed in the right panel of Fig. \ref{fig:regime-II-vary-mu-lamdm2dm1} and can produce the FIMP DM with the correct DM relic density. Therefore, our study on the two-component DM scenario in the regime III is new and interesting.
We provide analytical expressions for the decay and $2 \rightarrow 2$ contact annihilation processes in Appendix \ref{apdx:analyticFI}; see Eqs. \eqref{eqn:decay-contribution}--\eqref{eqn:scattY2-2}.

The Boltzmann equations associated with the WIMP and FIMP DM are given by
\begin{align}
\frac{d Y_{1}}{d x} &= 
-\frac{2 \pi^2}{45} \frac{M_{{\rm Pl}} M_{1}}{1.66 x^{2}}
\sqrt{g_{*}} \langle \sigma v \rangle_{\rm th}
\left( Y^{2}_{1} - Y^{{\rm eq}2}_{1} \right)  
\,, \label{eqn:regime-II-BE} \\ 
\frac{d Y_{2}}{d x} &=
\frac{2 M_{{\rm Pl}}}{1.66 M^2_{1}} 
\frac{x \sqrt{g_{*}(x)}}{g_{s}(x)}
\sum_{i = 1,2} \langle \Gamma_{h_{i} 
\rightarrow \phi^{\dagger}_{1} \phi_1} \rangle 
\left( Y^{{\rm eq}}_{h_i} - Y^{2}_{2} \right) 
+ \frac{4 \pi^2}{45} \frac{M_{{\rm Pl}} M_{1} \sqrt{g_{*}(x)}}{1.66 x^{2}}
\sum_{i,j \in {\rm SM}, \phi_1} \langle \sigma v \rangle_{ij} 
\left( Y^{{\rm eq}}_{i} Y^{{\rm eq}}_{j} - Y^{2}_{2}\right)  
\,.\nonumber
\end{align}
In the following, we solve the above Boltzmann equations and discuss the correlation between the model parameters by performing scans with the following range:
\begin{gather}
10^{-3} \leq  \theta \leq  10^{-1}\,, \qquad
10^{-3}  \leq \lambda_{1h},\lambda_{1H} \leq  10^{-1}\,, \qquad
10^{-12}  \leq \lambda_{2h},\lambda_{2H}, \lambda_{12} \leq  10^{-10}
\,, \nonumber \\
10^{-10}  \leq n_{\mu\tau} \leq  10^{-8}\,, \qquad
10^{-4}  \leq g_{\mu\tau} \leq  10^{-2} \,, \label{eqn:scanRange} \\
10^{-3}  \leq M_{Z_{\mu\tau}} [{\rm GeV}] \leq  1 \,, \qquad
200  \leq M_{h_2} [{\rm GeV}] \leq  1100 \,, \qquad
1  \leq M_{1,2} [{\rm GeV}] \leq  1000 \,.  \nonumber
\end{gather}
When performing the scans, we demand the total DM relic density to be in the range $0.01 \leq \Omega_{\rm DM} h^{2} \leq 0.12$. We stress that, when the sum of the WIMP and FIMP DM relic densities is smaller than $\Omega_{\rm DM} h^{2} = 0.12$, the rest of the amount can easily be obtained by suitably adjusting the $\mu$ parameter which we neglect at the moment. 

\begin{figure}[t!]
\centering
\includegraphics[angle=0,height=7cm,width=7cm]{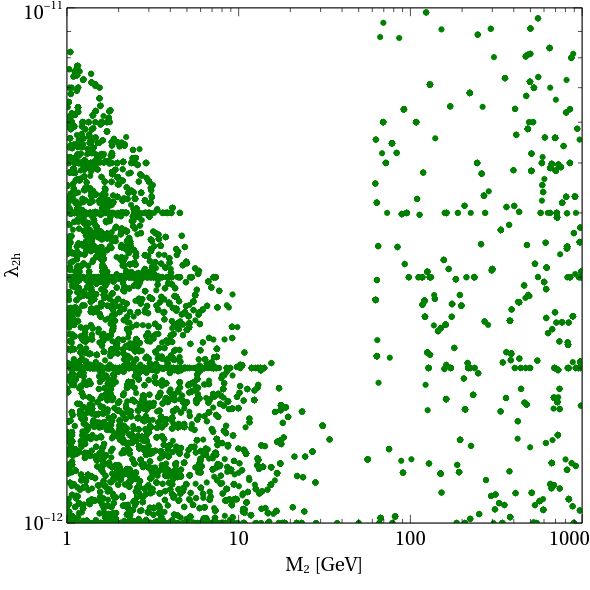}
\qquad
\includegraphics[angle=0,height=7cm,width=7cm]{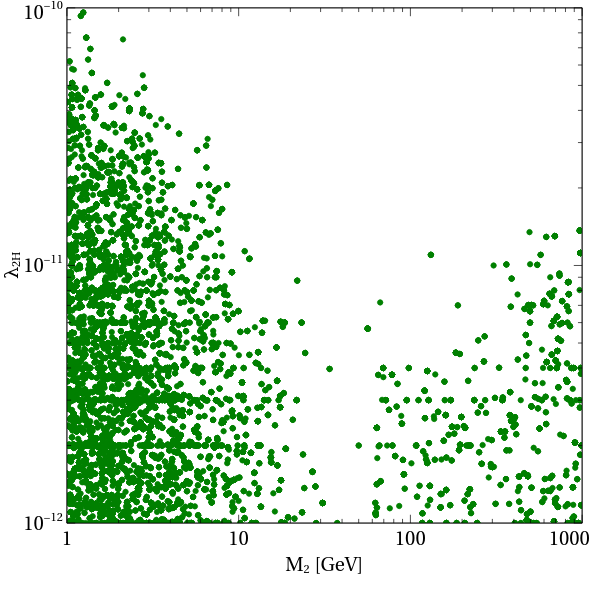}
\caption{Allowed parameter space in the $M_{2}$ -- $\lambda_{2h}$ plane ({\it left}) and in the $M_{2}$ -- $\lambda_{2H}$ plane ({\it right}) with the total DM relic density in the range $0.01 \leq \Omega_{\rm DM} h^{2} \leq 0.12$. The parameter ranges used in the scan are given in Eq. \eqref{eqn:scanRange}.}
\label{fig:regime-III-vary-mfimp-lambdahi}
\end{figure}

The allowed parameter regions in the $M_{2}$ -- $\lambda_{2h}$ and $M_{2}$ -- $\lambda_{2H}$ planes after imposing $0.01 \leq \Omega_{\rm DM} h^{2} \leq 0.12$ are shown in Fig. \ref{fig:regime-III-vary-mfimp-lambdahi}.
From the left panel, we see a sharp correlation between the FIMP DM mass $M_{2}$ and the coupling $\lambda_{2h}$ for $M_{2} \lesssim 62$ GeV. The sharp correlation may be understood as follows. For $M_{2} \lesssim 62$ GeV, the DM can be produced from the SM Higgs decay. The decay mode $h_{1} \rightarrow \phi^{\dagger}_{2} \phi_{2}$ is proportional to $\lambda_{1h}$ and the phase-space factor $\sqrt{1 - 4M^2_{2}/M_{h_1}^2}$. Moreover, the DM relic density is proportional to the DM mass as well. 
Thus, as the DM mass increases, the $\lambda_{2h}$ coupling needs to be decreased in order to obtain the correct DM relic density.
For DM mass in the range $50$ -- $60$ GeV, we have the phase-space suppression. Hence, in order to get DM in this range, we need a larger value of $\lambda_{2h}$.
For $M_{2} \gtrsim 62.5$ GeV, we do not have the decay channel of the SM Higgs into the FIMP DM, and annihilation processes take over. We thus do not have a sharp correlation in the large $M_2$ region. 

From the right panel of Fig. \ref{fig:regime-III-vary-mfimp-lambdahi}, we also observe a similar kind of behaviour between $M_{2}$ and $\lambda_{2H}$. However, for $M_{2} \gtrsim 62.5$ GeV, the $\lambda_{2H}$ parameter cannot be arbitrarily large as the $h_2$ decay mode is present. Moreover, we see that $\lambda_{2H}$ may become as large as $10^{-10}$ whereas $\lambda_{2h}$ can go only up to $10^{-11}$. The reason for this is that we varied $M_{h_2}$ up to 1.1 TeV and that the FIMP DM relic density through decay is proportional to $\lambda^2_{2H}/M_{h_2}$. Therefore, $\lambda_{2H}$ may become larger as $M_{h_2}$ takes a larger value, which is impossible for the SM Higgs case. 

\begin{figure}[t!]
\centering
\includegraphics[angle=0,height=7cm,width=7cm]{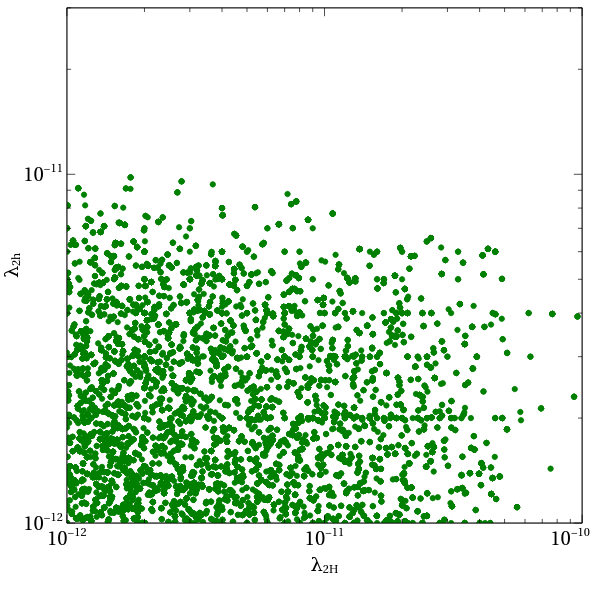}
\qquad
\includegraphics[angle=0,height=7cm,width=7cm]{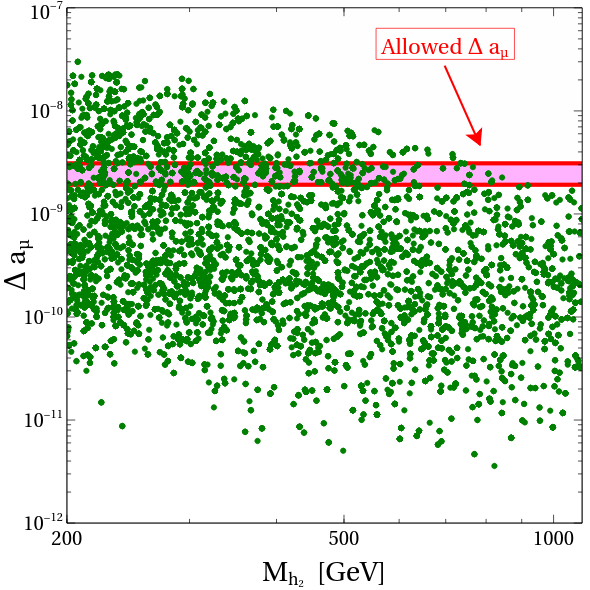}
\caption{
Allowed parameter space in the $\lambda_{2h}$ -- $\lambda_{2H}$ plane ({\it left}) and in the $M_{h_2}$ -- $\Delta a_{\mu}$ plane ({\it right}) with the total DM relic density in the range $0.01 \leq \Omega_{\rm DM} h^{2} \leq 0.12$.
The muon $g-2$ tension may be relieved in the magenta-coloured region.
The parameter ranges used in the scan are given in Eq. \eqref{eqn:scanRange}.
} 
\label{fig:regime-III-vary-delta-mu-lambdahi}
\end{figure}

The left panel of Fig. \ref{fig:regime-III-vary-delta-mu-lambdahi} shows the allowed parameter region in the $\lambda_{2h}$ -- $\lambda_{2H}$ plane. Since we have considered both the WIMP and FIMP DM in the 
DM relic density bound, it is hard to bound the quartic couplings from below as there will always be a contribution from the WIMP DM. However, we may obtain an upper bound on $\lambda_{2h}$ and $\lambda_{2H}$ above which the DM is overproduced. We find the upper limits as $\lambda_{2h} \lesssim 10^{-11}$ and $\lambda_{2H} \lesssim 10^{-10}$ for the choice of model parameters used in the scan \eqref{eqn:scanRange}.

The right panel of Fig. \ref{fig:regime-III-vary-delta-mu-lambdahi} shows the allowed parameter region in the $M_{h_2}$ -- $\Delta a_{\mu}$ plane.
The magenta-coloured region corresponds to the correct experimental range of muon $g-2$. The parameter space above the allowed magenta-coloured band is ruled out, and the points below the band demand additional positive contributions in $(g-2)_{\mu}$ to match the experimental range.
One may see a correlation between $\Delta a_{\mu}$ and $M_{h_2}$. For a larger value of $M_{h_2}$, we get a lower value of $\Delta a_{\mu}$. This is due to the fact that a higher value of $M_{h_2}$ indicates a higher VEV of the BSM Higgs, $v_{\mu\tau}$. Since $v_{\mu\tau} = M_{Z_{\mu\tau}}/g_{\mu\tau}$, a higher VEV implies a lower value of $g_{\mu\tau}$ which reduces the $\Delta a_{\mu}$ contribution.

\begin{figure}[t!]
\centering
\includegraphics[angle=0,height=7cm,width=7cm]{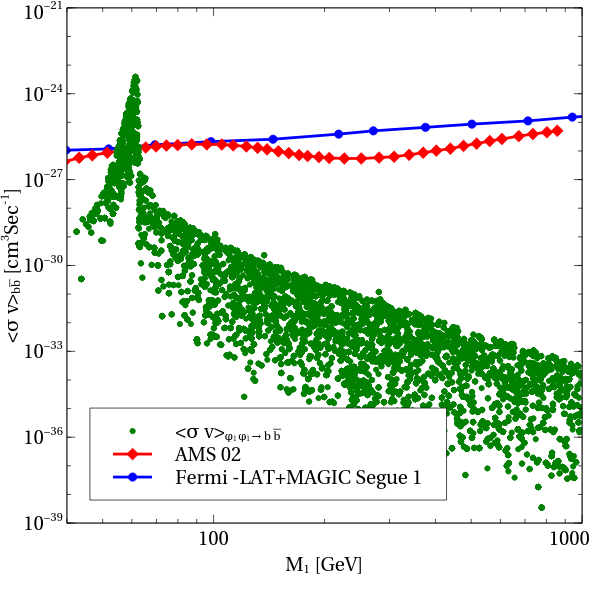}\;\qquad
\includegraphics[angle=0,height=7cm,width=7cm]{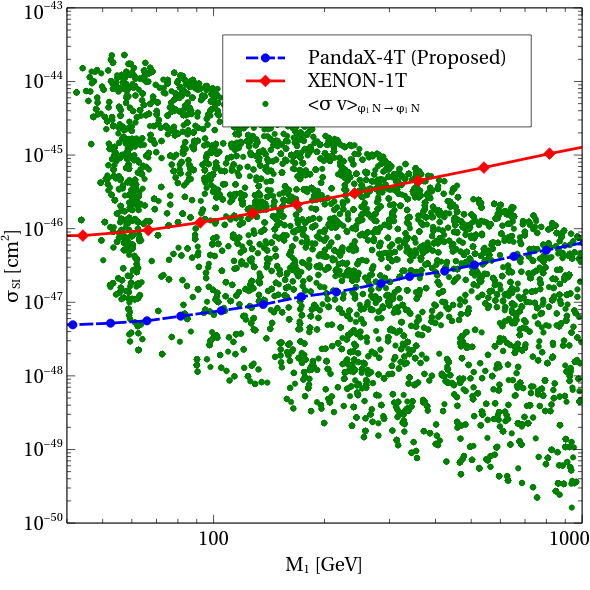}
\caption{
{\it Left}: Results of the scans in the $M_{1}$ -- $\langle \sigma v \rangle_{b \bar b}$ plane. The points above the red or blue lines are ruled out by indirect detection experiments \cite{MAGIC:2016xys,Reinert:2017aga}.
{\it Right}: Results of the scans in the $M_{1}$ -- $\sigma_{\rm SI}$ plane. The points above the red line is ruled out from the direct detection experiment Xenon-1T \cite{XENON:2018voc}. The blue line indicates the sensitivity of the future direct detection experiment PandaX \cite{PandaX:2018wtu}.
The parameter ranges used in the scan are given in Eq. \eqref{eqn:scanRange}.
}
\label{fig:regime-III-vary-mwimp-sigv-sigsi}
\end{figure}

In Fig. \ref{fig:regime-III-vary-mwimp-sigv-sigsi}, we present the indirect and direct detection bounds on the mass of the WIMP DM. In the left panel, the AMS 02 indirect detection bound coming from the WIMP DM annihilation to $b\bar{b}$ \cite{MAGIC:2016xys,Reinert:2017aga} is indicated with a red line. We see a sharp rise around $M_{1} \simeq 62$ GeV which corresponds to the SM Higgs resonance region.
A part of the region is ruled out by the indirect detection bound and the rest of the region is to be probed shortly by different ongoing indirect detection experiments \cite{MAGIC:2016xys}.
In the right panel, the spin-independent direct detection XENON-1T bound \cite{XENON:2018voc} on the WIMP DM is shown.
One may see from Fig. \ref{fig:regime-III-vary-mwimp-sigv-sigsi} that a part of the $M_{1} \leq 500$ GeV region is already ruled out from the direct detection experiments. The rest of the region will be explored in the future by different proposed experiments like Darwin \cite{DARWIN:2016hyl} and PandaX \cite{PandaX:2018wtu}.

%%%%%%%%%%%%%%%%%%%%%%%%%%%%%%%%%%%
\section{Gravitational Waves from Phase Transition}
\label{sec:FOPTGW}
%%%%%%%%%%%%%%%%%%%%%%%%%%%%%%%%%%%
The presence of the extra scalar fields in addition to the SM Higgs field not only makes the phenomenology of DM much richer, but it also makes the evolution dynamics of the vacuum state non-trivial and may lead to a FOPT in the early universe as opposed to the SM case whose phase transition is of the cross-over type \cite{Kajantie:1996mn}. See, {\it e.g.}, Ref.  \cite{Hindmarsh:2020hop} for a recent review on the FOPT.
As a consequence, stochastic GW signals may be emitted. The produced stochastic GW signals have a potential to be detected by future GW experiments such as LISA \cite{Baker:2019nia}, DECIGO \cite{Seto:2001qf}, and BBO \cite{Corbin:2005ny}, and this possibility gives a complementary detection signal to the standard (in-)direct detection and collider searches. In this section, we examine stochastic GW signals from a FOPT in our model and compare them with the sensitivity curves of future GW experiments. Furthermore, we present benchmark points that explain the muon $g-2$ tension, neutrino masses, and correct DM relic density, while producing strong GW signals that are within the detectability of Ultimate-DECIGO.

We closely follow Ref. \cite{Caprini:2015zlo} to estimate the stochastic GW signal from a FOPT. The three main sources of the GWs produced by a FOPT include the collision of bubble walls, the sound wave in the plasma, and the magneto-hydrodynamic turbulence in the plasma, and thus,
\begin{align}
\Omega_{\rm GW}h^{2}  \simeq 
\Omega_{\rm col}h^{2} 
+\Omega_{\rm sw}h^{2}
+\Omega_{\rm turb}h^{2}
\,,
\end{align}
where 
\begin{align}\label{eqn:Oh2col}
\Omega_{\rm col} h^{2} =
1.67 \times 10^{-5}
\left(\frac{H_{*}}{\beta}\right)^{2}
\left(\frac{\kappa_{\phi} \alpha}{1+\alpha}\right)^{2}
\left(\frac{100}{g_{*}}\right)^{\frac{1}{3}}
\left(\frac{0.11 v_{w}^{3}}{0.42+v_{w}^{2}}\right)
\left( \frac{3.8\left(f / f_{\rm col}\right)^{2.8}}{1+2.8\left(f / f_{\rm col}\right)^{3.8}} \right)
\,,
\end{align}
\begin{align}
\Omega_{\rm sw} h^{2} =
2.65 \times 10^{-6}
\left(\frac{H_{*}}{\beta}\right)
\left(\frac{\kappa_{v} \alpha}{1+\alpha}\right)^{2}
\left(\frac{100}{g_{*}}\right)^{\frac{1}{3}} v_{w} 
\left(f / f_{\rm sw}\right)^{3}
\left(\frac{7}{4+3\left(f / f_{\rm sw}\right)^{2}}\right)^{\frac{7}{2}}
\,,
\end{align}
and
\begin{align}
\Omega_{\rm turb} h^{2} =
3.35 \times 10^{-4}
\left(\frac{H_{*}}{\beta}\right)
\left(\frac{\kappa_{\rm turb} \alpha}{1+\alpha}\right)^{\frac{3}{2}}
\left(\frac{100}{g_{*}}\right)^{\frac{1}{3}}   
\left( \frac{v_{w} \left(f / f_{\rm turb}\right)^{3}}{\left[1+\left(f / f_{\rm turb}\right)\right]^{\frac{11}{3}}\left(1+8 \pi f / h_{*}\right)} \right)
\,,
\end{align}
with
\begin{align}
h_{*}=1.65 \times 10^{-5} \,{\rm Hz}
\left(\frac{T_{*}}{100 {\rm GeV}}\right)
\left(\frac{g_{*}}{100}\right)^{\frac{1}{6}}
\,.
\end{align}
The expressions for $f_{\rm col}$, $f_{\rm sw}$, and $f_{\rm turb}$ are given as follows:
\begin{align}
f_{\rm col} =
1.65 \times 10^{-5} \, {\rm Hz}
\left(\frac{0.62}{1.8-0.1 v_{w}+v_{w}^{2}}\right)
\left(\frac{\beta}{H_{*}}\right)
\left(\frac{T_{*}}{100 {\rm GeV}}\right)
\left(\frac{g_{*}}{100}\right)^{\frac{1}{6}}
\,,
\end{align}
\begin{align}
f_{\rm sw} =
1.9 \times 10^{-5} \, {\rm Hz} 
\left(\frac{1}{v_{w}}\right)
\left(\frac{\beta}{H_{*}}\right)
\left(\frac{T_{*}}{100 {\rm GeV}}\right)
\left(\frac{g_{*}}{100}\right)^{\frac{1}{6}}
\,,
\end{align}
and 
\begin{align}\label{eqn:fturb}
f_{\rm turb} =
2.7 \times 10^{-5} \, {\rm Hz} 
\left(\frac{1}{v_{w}}\right)
\left(\frac{\beta}{H_{*}}\right)
\left(\frac{T_{*}}{100 {\rm GeV}}\right)
\left(\frac{g_{*}}{100}\right)^{\frac{1}{6}}
\,.
\end{align}
Here, $g_{*}$ is the number of effective degrees of freedom at $T=T_*$.
For the bubble wall velocity $v_w$, we use \cite{Steinhardt:1981ct}
\begin{align}
v_{w}=\frac{\sqrt{1 / 3}+\sqrt{\alpha^{2}+2 \alpha / 3}}{1+\alpha}
\,,
\end{align}
and we adopt \cite{Kamionkowski:1993fg}
\begin{align}
\kappa=
\frac{0.715 \alpha+(4 / 27) \sqrt{3 \alpha / 2}}{1+0.715 \alpha}
\,, \qquad 
\kappa_{v}=
\frac{\alpha}{0.73+0.083 \sqrt{\alpha}+\alpha}
\,,\qquad
\kappa_{\rm turb} = 0.1 \kappa_{v}
\,,
\end{align}

In estimating the sound-wave contribution to the GW signal, we have ignored the possible suppression factor associated with the lifetime of the sound-wave source.\footnote{We thank the anonymous referee for pointing out this.} The suppression factor may be estimated as \cite{Ellis:2018mja,Ellis:2019oqb,Ellis:2020awk,Guo:2020grp}
\begin{align}\label{eqn:swSF}
\mathcal{S}_{\rm sw} \simeq 1.81 \times {\rm min}
\left\{
1, \frac{2(8\pi)^{1/3}}{\sqrt{3}}v_w\left(\frac{H_*}{\beta}\right)
\sqrt{\frac{1+\alpha}{\kappa_v \alpha}}
\right\}\,,
\end{align}
which corresponds to $\mathcal{O}\left(10^{-2}-10^{-1}\right)$ for the benchmark points (BPs) presented in Table \ref{tab:GWbenchamarkpoints}.
It is also important to note that such a suppression may be followed by a possible enhancement in the turbulence contribution to the GW signal \cite{Ellis:2018mja}. The precise determination requires dedicated and sophisticated numerical simulations which go beyond the scope of the present work.

From Eqs. \eqref{eqn:Oh2col} -- \eqref{eqn:fturb}, one may see that the key parameters that control the GW signal are $\alpha$, $\beta/H_*$, and $T_n$, where
\begin{align}
\alpha=\frac{\rho_{\rm vac}}{\rho_{\rm rad}^{*}}
\,,\qquad
\frac{\beta}{H_{*}}=
T_{*} \frac{d S_{\rm E}}{d T} \bigg\vert_{T_{*}}
\,,
\end{align}
with $S_{\rm E}$ being the Euclidean action of a bubble and $\rho_{\text{vac}}$ the energy density released during the FOPT.
We note that $\rho_{\rm rad}^{*}=g_{*} \pi^{2} T_{*}^{4} / 30$.
Throughout the section, we take $T_*$ to be the nucleation temperature $T_n$, {\it i.e.}, $T_* = T_n$.

To understand the dynamics of the FOPT, we use the one-loop effective potential,\footnote{
For a gauge dependence issue, readers may refer to Refs. \cite{Nielsen:1975fs,Fukuda:1975di,Patel:2011th,Chiang:2017zbz,Croon:2020cgk}.
}
\begin{align}
V_{\rm eff}^{\rm 1-loop} = V^{\rm tree} 
+ V^{\rm 1-loop}_{{\rm eff},T=0}
+ V^{\rm 1-loop}_{{\rm eff},T\neq0}
\,.
\end{align}
Here, $V^{\rm tree}$ is the tree-level potential and $V^{\rm 1-loop}_{{\rm eff},T=0}$ is the zero-temperature one-loop Coleman-Weinberg contribution \cite{Coleman:1973jx} which, in the $\overline{\rm MS}$ scheme, is given by
\begin{align}
V^{\rm 1-loop}_{{\rm eff},T=0} = 
\pm \sum_{i} n_i \frac{M_{i}^{4}}{64 \pi^{2}} 
\left[\ln \frac{M_{i}^{2}}{\Lambda^{2}}-c_{i}\right]
\,,
\end{align}
where $\Lambda$ is the renormalisation scale which we take to be $\Lambda^2 = (v^2+v_{\mu\tau}^2)/2$, $n_i$ is the number of degrees of freedom of the particle with field-dependent mass $M_i$, the constants $c_i$ are $1/2$ ($3/2$) for transverse gauge bosons (all other particles), and $+$ ($-$) is for bosons (fermions). 
The last correction, $V^{\rm 1-loop}_{{\rm eff},T\neq0}$, is the finite-temperature one-loop correction given by \cite{Dolan:1973qd}
\begin{align}
V^{\rm 1-loop}_{{\rm eff},T\neq0} =
\sum_{i} \frac{T^{4}}{2 \pi^{2}} n_{i} I_{\pm}
\left(\frac{m_{i}^{2}}{T^{2}}\right)
\,,
\end{align}
with
\begin{align}
I_{\pm}(x)=\pm \int_{0}^{\infty} dy \, y^{2} \ln 
\left(1 \mp e^{-\sqrt{y^{2}+x}}\right)
\,,
\end{align}
where $+$ ($-$) is for fermions (bosons).
To take into account the re-summed ring diagrams, we replace the field-dependent masses as
\begin{align}
M_{i}^{2} \rightarrow \widetilde{M}_{i}^{2}=M_{i}^{2}+\Pi_{i}(T)
\,,
\end{align}
where $\Pi_{i}(T)$ are the thermal masses \cite{Carrington:1991hz}. For the scalars in our model these corrections are
\begin{align}
\Pi_h &= \frac{T^2}{48}\left(
3g_1^2 + 9g_2^2 + 12y_t^2 + 12\lambda_h + 2\lambda_{hH} + 4\lambda_{1h} + 4\lambda_{2h}
\right)\,,\\
\Pi_H &= \frac{T^2}{24}\left(
6g_{\mu\tau}^2 + 6\lambda_{H} + \lambda_{hH} + 2\lambda_{1H} + 2\lambda_{2H} + 4h_{e\tau}^2
\right)\,,\\
\Pi_{\varphi_1} &= \Pi_{\eta_1} = \frac{T^2}{24}\left(
8\lambda_1 + 2\lambda_{12} + \lambda_{1h} + \lambda_{1H}
\right)\,,\\
\Pi_{\varphi_2} &= \Pi_{\eta_2} = \frac{T^2}{24}\left(
8\lambda_2 + 2\lambda_{12} + \lambda_{2h} + \lambda_{2H}
\right)\,,
\end{align}
where $\varphi_{1,2}$ ($\eta_{1,2}$) are the real (imaginary) components of the DM candidate $\phi_{1,2}$, and for the gauge boson, for which only the longitudinal mode receive corrections,
\begin{align}
\Pi_{W_L^{1,2,3}} = \frac{11}{6}g_2^2T^2 \,,\quad
\Pi_{B_L} = \frac{11}{6}g_1^2T^2 \,,\quad
\Pi_{Z_{\mu\tau L}} = \frac{1}{3}g_{\mu\tau}^2 T^2\,.
\end{align}
Fermions do not receive any corrections.

In order to estimate the $\alpha$ and $\beta/H_*$ parameters in our model, we performed a numerical analysis by using a modified version of \textsc{CosmoTransitions} \cite{Wainwright:2011kj} together with the mass spectra given above.
We restrict our focus on the case where only the SM and the BSM Higgses develop VEVs, taking zero VEVs for the DM candidates $\phi_1$ and $\phi_2$ throughout the temperature evolution of the system. 
We work with the following input parameters:
\begin{gather}
v_{\mu\tau} \,,\quad
M_{h_2} \,,\quad
\theta \,,\quad
M_{ee} \,,\quad
M_1 \,,\quad
M_2 \,,\quad 
\mu \,,\quad
g_{\mu\tau} \,,\quad
h_{e\tau} \,,\nonumber\\
\lambda_{1} \,,\quad
\lambda_{2} \,,\quad
\lambda_{1h} \,,\quad
\lambda_{1H} \,,\quad
\lambda_{12} \,,\quad
\lambda_{2h} \,,\quad
\lambda_{2H} \,,
\end{gather}
with the assumptions $M_{ee} = M_{\mu\tau}$ and $h_{e\tau} = h_{e\mu}$ which allow us to analytically diagonalise the RH neutrino mass matrix as we discussed in Sec. \ref{sec:model}. In the following, we take $h_{e\tau} = \sqrt{2}M_{ee}/v_{\mu\tau}$.
We impose the vacuum stability conditions,
\begin{align}
\lambda_h > 0\,, \quad 
\lambda_H > 0\,, \quad
4\lambda_h \lambda_H - \lambda_{hH}^2 \geq 0 \,,
\end{align}
as well as the perturbativity and unitarity bounds,
\begin{align}
\left|\lambda_{h}\right|<4 \pi, \quad\left|\lambda_{H}\right|<4 \pi, \quad\left|\lambda_{h H}\right|<8 \pi, \quad 3 \lambda_{h}+2 \lambda_{H}+\sqrt{\left(3 \lambda_{h}-2 \lambda_{H}\right)^{2}+2 \lambda_{h H}^{2}}<8 \pi
\,.
\end{align}
We focus on the following range of the parameters:
\begin{gather}
20 \leq v_{\mu\tau} \, [{\rm GeV}] \leq 250
\,,\quad
130 \leq M_{h_2} \, [{\rm GeV}] \leq 1000
\,,\quad
0 \leq \theta \leq 0.4
\,,\nonumber\\
1 \leq M_{ee} \, [{\rm GeV}] \leq 100
\,,\quad
0 \leq \lambda_{1h} \leq 0.5
\,,\quad
0 \leq \lambda_{1H} \leq 0.5
\,,\quad
10^{-12} \leq \lambda_{2h} \leq 10^{-10}
\,,\\
100 \leq M_1 \, [{\rm GeV}] \leq 300
\,,\quad
50 \leq M_2 \, [{\rm GeV}] \leq 250
\,.
\end{gather}
while fixing the other parameters as follows:
\begin{align}
\lambda_1 = \lambda_2 = 0.1
\,,\quad
\lambda_{2H} = \lambda_{12} = 6 \times 10^{-12}
\,,\quad
g_{\mu \tau} = 6 \times 10^{-4}
\,,\quad
\mu = 10^{-7}
\,.
\end{align}
We note that the upper bound of the mixing angle, $\theta = 0.4$, is chosen by considering the LHC constraints on the $hVV$ couplings coming from the measurements of the Higgs decay into gauge bosons \cite{ATLAS:2016neq}. We observe FOPTs and its associated GW signals for a wide range of the parameter values, including the mixing angle. To show that GWs can be accompanied with both small and large values of the mixing angle $\theta$, we present two BPs with a large value of $\theta$ and two BPs with a small value of $\theta$. 

\begin{figure}[t!]
\centering
\includegraphics[scale=0.9]{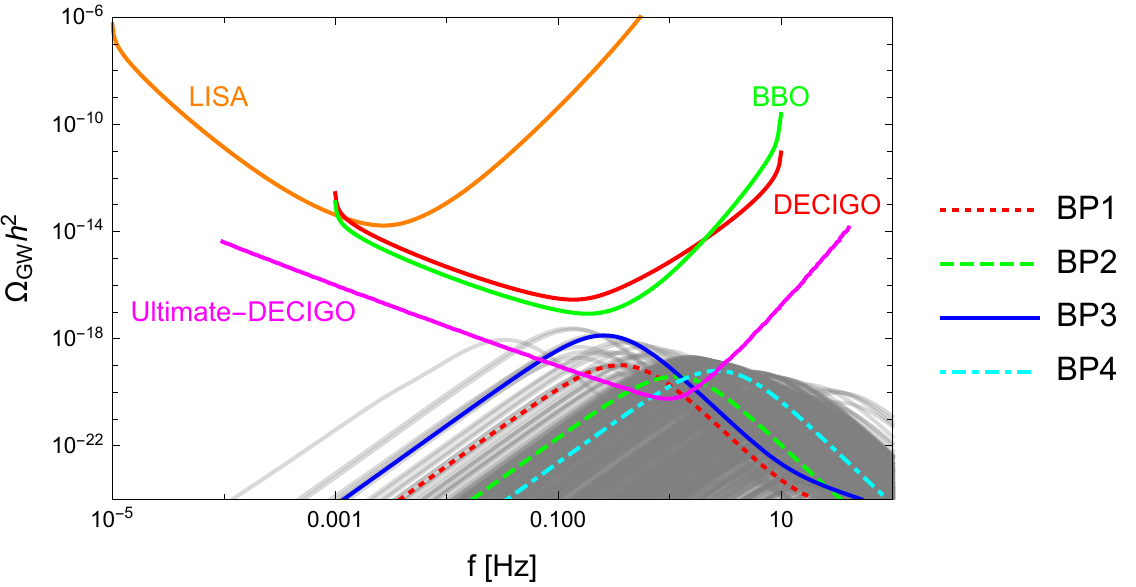}
\caption{GW spectrum from the FOPT together with the sensitivity curves of future GW experiments. The predicted GW signals span in the frequency range $0.01 \, {\rm Hz} \lesssim f \lesssim 100 \, {\rm Hz}$ with the magnitude as large as $\Omega_{\rm GW}h^2 \simeq 10^{-18}$. While the strengths of the signals are below the sensitivity curves of BBO and DECIGO, they are well within the reach of detectability of Ultimate-DECIGO. The data for the sensitivity curves of LISA, BBO, and DECIGO are obtained from Ref. \cite{Schmitz:2020syl}, and for the data for the sensitivity curve of Ultimate-DECIGO, we used Ref. \cite{Ringwald:2020vei}.
Earlier work on the sensitivity curves includes \textit{e.g.} Refs. \cite{Larson:1999we,Kudoh:2005as,Yagi:2011wg,Thrane:2013oya,Moore:2014lga,Kuroyanagi:2014qza,Saikawa:2018rcs,Robson:2018ifk}.
The red dotted, green dashed, blue solid, and cyan dot-dahsed lines correspond to our four benchmark points 1, 2, 3, and 4, respectively, that are summarised in Table \ref{tab:GWbenchamarkpoints}.
}
\label{fig:PTGW-plot}
\end{figure}

In Fig. \ref{fig:PTGW-plot}, we show the associated GW signals together with the sensitivity curves of future GW experiments. We select four BPs and present the results in Table \ref{tab:GWbenchamarkpoints}.
From Fig. \ref{fig:PTGW-plot}, we see that all of our four BPs, and many other signals, are well within the reach of detectability of Ultimate-DECIGO, while their signal strengths are below the sensitivity curves of BBO and DECIGO. 
Taking into account the suppression factor \eqref{eqn:swSF}, we see that some of the GW signals for the chosen BPs fall below the sensitivity curve of the Ultimate-DECIGO, while some stay marginally within the sensitivity curve. However, we stress that the conclusion that the GW signals associated with the FOPT within our model, which simultaneously accounts for the muon $g-2$ tension, neutrino masses, and two-component DM scenarios, are within the reach of the Ultimate-DECIGO sensitivity curve remains intact.

One may see from Table \ref{tab:GWbenchamarkpoints} that, in the parameter space that can solve the muon $g-2$ tension, generate the neutrinos masses, and produce the correct DM relic density, GWs are also expected whose signals are strong enough to be seen by Ultimate-DECIGO. The presented four BPs clearly showcase that the muon $g-2$, neutrino masses, and two-component DM scenarios are accounted for in our model which, at the same time, predicts stochastic GWs associated with the FOPT in a single unified framework.

\begin{table}[t!]
\centering
\begin{tabular}{|c|c|c|c|c|c|c|c|c|c|c|c|c|c|c|}
\hline \hline
BP & $v_{\mu\tau}$ & $M_{h_2}$ &
$M_{ee}$ & 
$M_1$ & $M_2$ &
$\theta$ & 
$\lambda_{2h}$  &
$\lambda_{1h}$ & $\lambda_{1H}$ &
$\alpha$ & $\beta/H$ & $T_n$ &
$\frac{\Omega_{\rm DM}^1}{\Omega_{\rm DM}}$ &
$\frac{\Omega_{\rm DM}^2}{\Omega_{\rm DM}}$ 
\\ \hline
1 & 80.14& 408.06 & 98.45 & 250.31 & 169.75 & 0.388 & 
$6.0 \times 10^{-12}$ & 
0.1 & 0.1 &
0.0036 & 4994.4 & 235.0 &
0.64 & 0.36 
\\ \hline
2 & 81.69& 415.34 & 99.83 & 244.57 & 99.39 & 0.387 &
$4.5 \times 10^{-12}$ & 
0.1 & 0.1 &
0.0037 & 15293.2 & 238.8 &
0.58 & 0.42
\\ \hline
3 & 81.66& 398.97 & 98.19 & 210.0 & 209.9 & 0.002 &
$4.5 \times 10^{-12}$ &
0.1 & 0.1 &
0.0068 & 4884.0 & 178.7 &
0.79 & 0.21 
\\ \hline
4 & 83.28& 366.09 & 67.07 & 249.51 & 119.54 & 0.034 & 
$1.7 \times 10^{-11}$ &
0.289 & 0.228 &
0.0056 & 47146.8 & 189.2 &
0.15 & 0.85
\\
\hline \hline
\end{tabular}
\caption{Four BPs. Values of the mass-dimensionful parameters are given in units of GeV. We present the model parameters, the GW-related parameters, $\alpha$, $\beta/H$, and $T_n$, and the DM relic densities $\Omega_{\rm DM}^{1,2}$. For all of these four BPs, the muon $g-2$ tension and the neutrino masses can be accounted for. The GW signals corresponding to the four BPs are highlighted in Fig. \ref{fig:PTGW-plot}.
The other parameters are chosen as $\lambda_{2H} = 6 \times 10^{-12}$, $\mu = 10^{-7}$, $\lambda_{12} = 6 \times 10^{-12}$, $g_{\mu\tau} = 6 \times 10^{-4}$, and $n_{\mu\tau} = 10^{-8}$. The neutrino masses can be obtained for the choice of parameters as studied in detail in Ref. \cite{Biswas:2016yan}, and $(g-2)_{\mu}$ will also be obtained in the correct ballpark value as shown in Fig. \ref{fig:muong2}. In determining the WIMP and FIMP DM individual contribution, we have considered the total DM relic density of $\Omega_{\rm DM} h^{2} = 0.12$.
}
\label{tab:GWbenchamarkpoints}
\end{table}
%%

%%%%%%%%%%%%%%%%%%%%%%%%%%%%%%%%%%%
\section{Conclusion}
\label{sec:conc}
%%%%%%%%%%%%%%%%%%%%%%%%%%%%%%%%%%%
In this paper, we studied an extension of the Standard Model that accounts for the dark matter, the muon $g-2$ tension, and the neutrino masses, in a single unified framework. We introduced three massive right-handed neutrinos which, through the type-I seesaw mechanism, provide a mass to the Standard Model neutrinos. We then extended the Standard Model by introducing two scalar fields that play the role of the dark matter. Finally, an extra $U(1)_{L_\mu - L_\tau}$ gauge symmetry is imposed, where the associated gauge boson $Z_{\mu\tau}$ alleviates the muon $g-2$ tension.

As the model we considered contains two Standard Model-singlet scalar dark matter candidates, $\phi_1$ and $\phi_2$, we examined the possibility of a single-component as well as two-component dark matter scenarios. Focusing on three different regimes, we showed how a single-component or two-component dark matter scenario can emerge by numerically solving the coupled Boltzmann equations. We found that, when the $\mu\phi_1^\dagger\phi_2^3$ term is not negligible, both the single- and two-component scenarios may be obtained, depending on the mass range. When the $\mu$ parameter is small or absent, we showed that a two-component dark matter scenario naturally arises without dependence on the mass range of the WIMP and FIMP DM. In the case of a two-component scenario, one component becomes the WIMP-type dark matter and the other component is the FIMP-type dark matter. We performed a numerical scan and presented viable parameter spaces which are compatible with the current experimental bounds such as the direct and indirect detections, relieving the muon $g-2$ tension at the same time.

The presence of the extra scalar fields not only makes the dark matter phenomenology richer. It also affects the evolution dynamics of the vacuum state. As opposed to the Standard Model case whose phase transition is of the cross-over type, a first-order phase transition may be realised in our model. Consequently, stochastic GW signals may be emitted. We investigated the parameter space where the first-order phase transition occurs and scrutinised the associated stochastic gravitational wave signals. Performing a numerical scan, we showed that the predicted gravitational waves are strong enough to be probed by future gravitational wave experiments such as Ultimate-DECIGO. 

We explicitly demonstrated that our model is capable of accommodating the three problems of the Standard Model, namely the dark matter, neutrino masses, and the muon $g-2$ tension, by presenting four benchmark points. The chosen four benchmark points give rise to the first-order phase transition, and consequently, we observe the associated gravitational wave signals. All of the four benchmark points are within the reach of detectability of Ultimate-DECIGO. Furthermore, the chosen benchmark points realise two-component dark matter scenarios.
We expect that the gravitational wave feature of our model may serve as a complementary detection signal to the standard (in-)direct detection and collider searches.

%%%%%%%%%%%%%%%%%%%%%%%%%%%%%%%%%%%
\section*{Acknowledgements}

The work of F.C. is supported by the European Union’s Horizon 2020 research and innovation programme under the Marie Skłodowska-Curie grant agreement No 860881-HIDDeN.
This work used the Scientific Compute Cluster at GWDG, the joint data center of Max Planck Society for the Advancement of Science (MPG) and University of G\"ottingen.

%%%%%%%%%%%%%%%%%%%%%%%%%%%%%%%%%%%

\appendix

%%%%%%%%%%%%%%%%%%%%%%%%%%%%%%%%%%%
\section{Analytical Expressions for the Freeze-in Dark Matter}
\label{apdx:analyticFI}
%%%%%%%%%%%%%%%%%%%%%%%%%%%%%%%%%%%

%%%%%%%%%%%%%%%%
\subsection{FIMP-WIMP interactions}
\label{apdx:FWint}
%%%%%%%%%%%%%%%%
We summarise different channels for the FIMP production, considering the interaction between the FIMP DM and the WIMP DM.

%%%%%%%%%%%%%%%%
\subsubsection{Exponential yield}
\label{apdx:exp-yield}
%%%%%%%%%%%%%%%%

In the regime where $M_1 < 3 M_2 $, the decay of the WIMP DM is kinetically forbidden. For the scattering process $ \phi_2 \phi_2 \leftrightarrow \phi_1 \phi_2$, the Boltzmann equation is given by
\begin{align}
\dot{n}_{\phi_2}+3 H n_{\phi_2} = 
\langle \sigma v \rangle n_{\phi_2} n^{{\rm eq}}_{\phi_1} \,,
\end{align}
to a good approximation.\footnote{The viability of the use of number densities in the Boltzmann equations is questioned and checked in, for example, Ref. \cite{Du:2021jcj} by considering the backreaction effects and solving the Boltzmann equations at the level of the phase-space distribution.} 
Since $\phi_2 $ is in a FIMP regime, {\it i.e.}, out of equilibrium, its number density is considerably low, and we can thus neglect the quadratic term in $n_{\phi_2}$. Notice that if the decay were allowed, the decay channel would become the dominant process as it may produce the total relic density of DM with a coupling orders of magnitude smaller than the scattering process as it is shown in Sec. \ref{sec:TCDM}. In terms of the yield $Y_{2}=n_{\phi_2}/S$, where $S$ is the entropy density, the Boltzmann equation can be re-written as
\begin{align}
\frac{dY_{2}}{dT} = - \frac{1}{HT}
\langle \sigma v  \rangle n^{{\rm eq}}_{\phi_1} Y_{2}
\,.
\end{align}
The solution has an exponential behaviour,
\begin{align}
Y_{2} = Y_0 \exp \left[ 
\int dT \, \frac{1}{HT} 
\langle \sigma v  \rangle 
n^{{\rm eq}}_{\phi_1 } 
\right]
\,.
\end{align}
The thermal average of cross section times velocity, $\langle \sigma v \rangle$, can be obtained by
\begin{align}
\langle \sigma v \rangle_{A\,B \rightarrow C\,D} = 
\frac{1}{8 M_{A}^{2} M_{B}^{2} 
K_{2}\left(M_{A}/T\right) K_{2}\left(M_{B}/T\right)} 
\int_{\left(M_{A}+M_{B}\right)^{2}}^{\infty} d s \, 
\frac{\sigma_{AB \rightarrow C D}}{\sqrt{s}} p_{AB} 
K_{1}\left(\frac{\sqrt{s}}{T}\right) 
\,,
\end{align}
where $M_{A,B}$ are the masses of $A$ and $B$, $s$ is the centre-of-mass energy, $p_{AB} = [s - ( M_{A} - M_{B} )^{2} ] [s - (M_{A} + M_{B} )^{2} ]$, and $K_{1,2}$ are the modified Bessel functions of the second kind.
In our model, the cross section is given by
\begin{align}
\sigma = \frac{|\mathcal{M}|^2}{32 \pi s} 
\sqrt{\frac{s(s-4 M_2^2)}{(s-(M_1+M_2)^2)(s-(M_1-M_2)^2)}}
\,,
\end{align}
where the matrix element is given by $|\mathcal{M}|^2=36\mu^2$.

%%%%%%%%%%%%%%%%
\subsubsection{Three-body decay}
%%%%%%%%%%%%%%%%
In the opposite regime, $M_1 > 3M_2$, the decay channel of the WIMP DM to the FIMP DM is open. The differential decay rate is given by
\begin{align}
d\Gamma = \frac{M_1 \pi}{32 (2 \pi)^5} 
|\mathcal{M}|^2 \,d x_3 \, d x_1 \, d\cos\theta \, d\phi,
\end{align}
where 
\begin{align}
&0 \leq \phi \leq 2 \pi \,,\quad
-1 \leq \cos\theta \leq 1 \,,\quad
2 \sqrt{c}  \leq x_{3} \leq 1+c-a-b-2 \sqrt{a} \sqrt{b} \,,\quad
x_{1}^{-} \leq x_{1} \leq x_{1}^{+} \,,\nonumber \\
&x_{1}^{\pm}= 1+a-b+c-x_{3}-\frac{1}{2}\left(2 c-x_{3}\right)\left(1+\frac{a-b}{1+c-x_{3}}\right)
\pm \frac{1}{2} y_{3} \sqrt{1-2 \frac{a+b}{1+c-x_{3}}+\frac{(a-b)^{2}}{\left(1+c-x_{3}\right)^{2}}}
\,,
\end{align}
with $a=b=c=M^2_2/M^2_1$ and $y_3 = \sqrt{x_3 -4c}$.
Therefore, we obtain the decay rate as follows:
\begin{align}\label{eqn:three-body-decay}
\Gamma = \frac{M_1}{256 \pi^3} 
|\mathcal{M}|^2 
\int^{1-3a}_{2 \sqrt{a}} dx_3 \sqrt{\frac{(x_3^2-4a)(1-3a-x_3)}{1+a-x_3}}
\,,
\end{align}
with $|\mathcal{M}|^2=36\mu^2$.

%%%%%%%%%%%%%%%%
\subsection{FIMP-SM/BSM Higgs interactions}
%%%%%%%%%%%%%%%%
We consider now different FIMP production channels through the interactions with the SM and BSM Higgses.
Rotating into the mass eigenstates and considering the electroweak broken phase, we have the following interaction:
\begin{align}
\mathcal{L}  &\supset  
2 v |\phi_2|^2 h_2 (\lambda_{2H} \cos{\theta} - \lambda_{2h} \sin{\theta})  
+ 2 v_{\mu \tau} |\phi_2|^2 h_1 (\lambda_{2H} \sin{\theta} + \lambda_{2h} \cos{\theta}) 
\nonumber \\ &\quad 
+ |\phi_2|^2 h_1 h_2 (\lambda_{2H} \cos{\theta} \sin{\theta} - \lambda_{2h} \cos{\theta} \sin{\theta}) 
+ \frac{1}{2} |\phi_2|^2 h_2^2 (\lambda_{2H} \cos^2{\theta} - \lambda_{2h} \sin^2{\theta}) 
\nonumber \\ & \quad
+ \frac{1}{2} |\phi_2|^2 h_1^2 (\lambda_{2H} \cos^2{\theta} - \lambda_{2h} \sin^2{\theta})
\\ &=
2 \lambda_{a}  v |\phi_2|^2 h_2 
+ 2 \lambda_{b}  v_{\mu \tau} |\phi_2|^2 h_1  
+ \lambda_{c} |\phi_2|^2 h_1 h_2 
+ \frac{1}{2}  \lambda_{d}   |\phi_2|^2 h_2^2 
+ \frac{1}{2}  \lambda_{e}  |\phi_2|^2 h_1^2 
\,,
\end{align}
where we have introduced new coupling constants as
\begin{align}
\lambda_{a} &\equiv
\lambda_{2H} \cos{\theta} - \lambda_{2h} \sin{\theta}
\,,\nonumber\\
\lambda_{b} &\equiv
\lambda_{2H} \sin{\theta} + \lambda_{2h} \cos{\theta}
\,,\nonumber\\
\lambda_{c} &\equiv
\lambda_{2H} \cos{\theta} \sin{\theta} - \lambda_{2h} \cos{\theta} \sin{\theta}
\,,\\
\lambda_{d} &\equiv
\lambda_{2H} \cos^2{\theta} - \lambda_{2h} \sin^2{\theta}
\,,\nonumber\\
\lambda_{e} &\equiv
\lambda_{2H} \cos^2{\theta} - \lambda_{2h} \sin^2{\theta}
\,.\nonumber
\end{align}
%%

%%%%%%%%%%%%%%%%
\subsubsection{Decay contribution}
%%%%%%%%%%%%%%%%

In the parameter space where $M_{h_1,h_2} > 2 M_2 $, the decays of the SM and the BSM Higgs fields are allowed, with the decay rates
\begin{align}
\Gamma_{\rm SM} = \frac{v_{\mu \tau}^2}{4 \pi M_{h_1}} \lambda_{b} \sqrt{M_{h_1}^2 - 4 M_2^2}
\,,\qquad
\Gamma_{\rm BSM} = \frac{v^2}{4 \pi M_{h_2}} \lambda_{a} \sqrt{M_{h_2}^2 - 4 M_2^2} 
\,.
\end{align}
Solving the Boltzmann equation, we obtain
\begin{align}
Y_2 &= 
\int dT \, \frac{M_{h_1}^2}{2 \pi^2 H S} 
K_1 \left(\frac{M_{h_1}}{T}\right) 
\Gamma_{\rm SM}
+ \int dT \, \frac{M_{h_2}^2}{2 \pi^2 H S}
K_1 \left(\frac{M_{h_2}}{T}\right) 
\Gamma_{\rm BSM}
\nonumber\\
&=
\frac{135M_{\rm Pl} }{1.66\cdot8 \pi^3 g_*^S \sqrt{g_*}} \left( \frac{ \Gamma_{\rm SM}}{M_{h_1}^2} + \frac{\Gamma_{\rm BSM}}{M_{h_2}^2}  \right)
\,.\label{eqn:decay-contribution}
\end{align}
We note that, when the decay channel is kinematically open, it dominates the production.

%%%%%%%%%%%%%%%%
\subsubsection{Scattering contribution}
%%%%%%%%%%%%%%%%
When the decay is inactive, the dominant contribution to the FIMP production comes from the scattering process. The yield in this case is given by 
\begin{align}\label{eqn:scattering-contribution}
\frac{d Y_2}{dT} = 
-\frac{1}{32 \pi^4 H S}  
\int_{4 M_{h_{12}}^{2}}^{\infty} ds \, 
\sigma\left(s-4 M_{h_{12}}^{2}\right) 
K_1 \left(\frac{\sqrt{s}}{T}\right)
\,,
\end{align}
where $M_{h_{12}}$ is $M_{h_1}$ ($M_{h_2}$) if the relevant interaction is governed by the coupling $\lambda_e$ ($\lambda_d$). 
In the $M_{h_1} \ll M_2$ limit, we get
\begin{align}\label{eqn:scattY2-1}
Y_2 = 
\frac{135 \lambda_{e}^2 M_{\rm Pl}}{1.66 \cdot 4096 \pi^4 g_*^S \sqrt{g_*}M_2^2} 
\,,
\end{align}
while in the $M_{h_2} \ll M_2$ limit, we find
\begin{align}\label{eqn:scattY2-2}
Y_2 = 
\frac{135 \lambda_{d}^2 M_{\rm Pl}}{1.66 \cdot 4096 \pi^4 g_*^S \sqrt{g_*}M_2^2} 
\,.
\end{align}
As $h_1$ represents the SM Higgs in our consideration, the first case, namely $M_{h_1} \ll M_2$, always holds to be the case.

%%%%%%%%%%%%%%%%%%%%%%%%%%%%%%%%
%%%%%%%%%%%%%%%%%%%%%%%%%%%%%%%%

\end{document}